\documentclass[a4paper,12pt]{article}
\usepackage{amsfonts}
\usepackage{latexsym}
\usepackage{amsmath}
\usepackage{amssymb}
\usepackage{amssymb}
\usepackage{slashed}
\usepackage{upgreek}
\usepackage{mathrsfs}
\usepackage{bbm}

\usepackage{relsize}
\usepackage{graphicx}
\usepackage{comment}

\newcommand{\newsection}{    
\setcounter{equation}{0}\section}
\def\appendix#1{\addtocounter{section}{1}\setcounter{equation}{0}
\renewcommand{\thesection}{\Alph{section}}
\section*{Appendix \thesection\protect\indent \parbox[t]{11.15cm}{#1}}
\addcontentsline{toc}{section}{Appendix \thesection\ \ \ #1}}

\def\bbe{{\bf{e}}}
\def\bbl{{\bf{\ell}}}
\font\mybb=msbm10 at 11pt

\def\bb#1{\hbox{\mybb#1}}

\def\bR {\bb{R}}

\def\bC {\bb{C}}

\def\gom{\Gamma\mkern-4.0mu \omega}
\def\gY{\Gamma\mkern-2.0mu Y}
\def\gF{\Gamma\mkern-4.0mu F}

\def\gH{\Gamma\mkern-4.0mu H}

\def\sY{\slashed {Y}}
\def\sgY{\slashed {\gY}}

\def\sF{\slashed {F}}
\def\sgF{\slashed {\gF}}

\def\sH{\slashed{H}}
\def\sgH{\slashed{\gH}}

\newcommand{\bea}{\begin{eqnarray}}
\newcommand{\eea}{\end{eqnarray}}

\usepackage[usenames,dvipsnames,svgnames,table]{xcolor}	
\usepackage[normalem]{ulem}				
\usepackage{microtype}
\usepackage[colorlinks, citecolor=blue, linkcolor=blue, urlcolor=Maroon, filecolor=Maroon, linktocpage=true]{hyperref} 

\begin{document}

\begin{center}
\vspace*{-1.0cm}
\begin{flushright}
\normalsize{\texttt{ZMP-HH/17-24}}\\
\end{flushright}

\vspace{2.0cm} {\Large \bf AdS$_4$ backgrounds with $N>16$ supersymmetries in $10$ and $11$ dimensions } \\[.2cm]

\vskip 2cm
A.~S.~Haupt$^{1,2}$, S.  Lautz$^3$ and  G.  Papadopoulos$^3$
\\
\vskip .6cm

\begin{small}
$^1$\textit{Department of Mathematics \\ and Center for Mathematical Physics,\\
University of Hamburg,\\ Bundesstr. 55, D-20146 Hamburg, Germany}
\end{small}\\*[.6cm]

\begin{small}
$^2$\textit{II. Institute for Theoretical Physics,\\ University of Hamburg,\\
Luruper Chaussee 149, D-22761 Hamburg, Germany}\\
\texttt{alexander.haupt@uni-hamburg.de}
\end{small}\\*[.6cm]

\begin{small}
$^3$\textit{Department of Mathematics, King's College London
\\
Strand, London WC2R 2LS, UK}\\
\texttt{sebastian.lautz@kcl.ac.uk}\\
\texttt{  george.papadopoulos@kcl.ac.uk}
\end{small}\\*[.6cm]

\end{center}

\vskip 2.5 cm

\begin{abstract}
\noindent
We explore  all warped   $AdS_4\times_w M^{D-4}$ backgrounds  with the most general allowed fluxes that preserve more than 16 supersymmetries  in $D=10$- and $11$-dimensional  supergravities.  After  imposing  the assumption that  either the internal space $M^{D-4}$
  is  compact without boundary or  the isometry algebra of the background decomposes into that of AdS$_4$ and that of  $M^{D-4}$,  we find that
there are no such backgrounds in IIB supergravity.  Similarly in IIA supergravity,  there is a unique such background with 24 supersymmetries locally isometric to
$AdS_4\times \mathbb{CP}^3$, and  in $D=11$ supergravity  all such backgrounds are locally isometric to the maximally supersymmetric
$AdS_4\times S^7$ solution.
	
\end{abstract}

\newpage

\tableofcontents


\section{Introduction}

AdS backgrounds in 10 and 11 dimensions that preserve $N$ supersymmetries with $N>16$
have found widespread applications both in supergravity compactifications
and in AdS/CFT correspondence, for reviews see \cite{duff, maldacena}
and references therein. One of the features of such  backgrounds in AdS/CFT \cite{maldacenab}
  is that the CFT R-symmetry group  acts transitively on the internal space of the solution and  this can be used
  to establish the dictionary between some of the operators of the CFT
and spacetime Kaluza-Klein fields \cite{witten}.  Therefore the question arises whether it is possible to find all such AdS solutions.
Despite the progress that has been made during the years, a complete description of all AdS solutions that
preserve $N>16$ supersymmetries remains an open problem.

Recently however,
there have been several developments which facilitate progress in this direction for a large class
of warped flux AdS solutions. In \cite{mads, iibads, iiaads},  the Killing spinor equations (KSEs) of supergravity theories  have been solved in all generality and the fractions of supersymmetry preserved  by all warped flux AdS backgrounds have been identified.  Furthermore  global analysis techniques have also been introduced
in the investigation of AdS backgrounds which can be used to  a priori impose properties like the compactness of the internal space and the smoothness of the fields.  Another key development is the proof of the
homogeneity theorem \cite{homogen} which for the special case of AdS backgrounds states  that all such
backgrounds that preserve $N>16$ supersymmetries are Lorentzian homogeneous spaces.

 So far it  is known that the warped flux   AdS$_n$, $n\geq 6$, backgrounds preserve either 16 or 32 supersymmetries
and those that preserve 32 supersymmetries have been classified in \cite{maxsusy}. In addition, it has been shown that there are no  $N>16$
 AdS$_5$ backgrounds in $D=11$ and (massive) IIA supergravities while in IIB supergravity all such backgrounds are locally isometric to the maximally
  supersymmetric AdS$_5\times S^5$ solution \cite{ads5clas}.  In particular the existence of a IIB AdS$_5$ solution that preserves 24 supersymmetries has been excluded.
 Moreover the $AdS_n\times M^{D-n}$  solutions with  $M^{D-n}$  a symmetric coset space  have been classified
in \cite{figueroaa, figueroab, Wulffa, Wulffb}. Furthermore heterotic supergravity does not admit  AdS  solutions that preserve more than 8 supersymmetries \cite{adshet}.

 The main task of this paper is to describe all warped  AdS$_4$ backgrounds that admit   the most general   fluxes  in 10 and 11 dimensions
and preserve more than 16 supersymmetries. It has been shown in  \cite{mads, iibads, iiaads} that such backgrounds preserve $4k$ supersymmetries. Therefore, we shall investigate
the backgrounds preserving 20, 24 and  28 as those with  32 supersymmetries have already been classified in \cite{maxsusy}.
In particular,  we find that
\begin{itemize}
\item  IIB and massive IIA supergravity do not admit  AdS$_4$ solutions with $N>16$ supersymmetries.
\item  Standard  IIA supergravity admits a unique  solution up to an overall scale preserving 24 supersymmetries   locally isometric to the AdS$_4\times \mathbb{CP}^3$
background of  \cite{nillsonpope}.
\item  All  AdS$_4$ solutions of 11-dimensional supergravity that preserve  $N>16$  supersymmetries are locally isometric to the maximally
supersymmetric AdS$_4\times S^7$ solution of  \cite{fr, duffpopemax}.
\end{itemize}

These results have been established under certain  assumptions\footnote{Some assumptions are necessary to exclude the possibility that a warped  AdS$_4$
 background is not locally isometric to  an AdS$_n$ background with $n>4$. This has been observed in \cite{strominger} and explored in the context of KSEs in \cite{desads}.}. We begin with a spacetime which is a warped product $AdS_4\times_w M^{D-4}$, for $D=10$ or $11$, and allow for all fluxes which are invariant  under  the isometries of AdS$_4$. Then
we shall assume  that
\begin{enumerate}
\item  either the solutions are smooth and  $M^{D-4}$ is compact without boundary
\item  or that the even part of the Killing  superalgebra of the background
decomposes as a direct sum $\mathfrak{so}(3,2)\oplus \mathfrak{t}_0$, where $\mathfrak{so}(3,2)$ is the Lie algebra of isometries of AdS$_4$ and $\mathfrak{t}_0$ is the Lie algebra of the isometries of $M^{D-4}$.
\end{enumerate}
It has been shown in \cite{superalgebra} that for all AdS backgrounds, the first assumption implies the second. In addition for $N>16$  AdS$_4$ backgrounds\footnote{In what follows,
we use ``$N>16$ AdS backgrounds'' instead of ``AdS backgrounds that preserve $N>16$ supersymmetries'' for short.}, the second assumption implies the first.   This is because  $\mathfrak{t}_0$ is the Lie algebra of a compact group and all internal
 spaces are compact without boundaries. Smoothness also follows as a consequence of
 considering only invariant solutions.

The proof of the main statement  of our paper is based first on the results of \cite{mads, iibads, iiaads} that the number of supersymmetries preserved by AdS$_4$ backgrounds are $4k$
 and so the solutions under consideration preserve  20, 24, 28 and 32 supersymmetries. Then the homogeneity theorem of \cite{homogen} implies that all such backgrounds
 are Lorentzian homogeneous spaces. Moreover, it has been shown in  \cite{superalgebra} under the assumptions mentioned above that the Killing
superalgebra of warped AdS$_4$ backgrounds that  preserve $N=4k$ supersymmetries  is isomorphic to $\mathfrak{osp}(N/4 \vert4)$, see also \cite{charles},  and that the even subalgebra $\mathfrak{osp}(N/4 \vert4)_0=\mathfrak{so}(3,2)\oplus \mathfrak{so}(N/4)$ acts  effectively on the spacetime with $\mathfrak{t}_0=\mathfrak{so}(N/4)$ acting on the internal space.
Thus together with the homogeneity theorem $\mathfrak{osp}(N/4 \vert4)_0$ acts both {\sl transitively} and {\sl effectively} on the spacetime.
Then we demonstrate in all cases that the warp factor $A$ is constant. As a result all $N>16$  AdS$_4$ backgrounds
are product spaces $AdS_4\times M^{D-4}$.  So the internal space $M^{D-4}$ is a homogeneous space, $M^{D-4}=G/H$, and $\mathfrak{Lie}\, G= \mathfrak{so}(N/4)$.
Therefore, we  have demonstrated the following,
\begin{itemize}
\item The internal spaces of AdS$_4$ backgrounds that preserve $N>16$ supersymmetries are homogeneous spaces that  admit
a transitive and effective action of a group $G$ with  $\mathfrak{Lie}\, G= \mathfrak{so}(N/4)$.
\end{itemize}
Having established this,
one can use the classification of   \cite{ Castellani:1983yg, klausthesis, niko6dim, niko7dim} to identify all the 6- and 7-dimensional homogeneous spaces that can occur as internal spaces
for $N>16$  AdS$_4$ backgrounds, see also tables\footnote{These tables list the simply connected homogeneous spaces. This suffices for our purpose because we are investigating the  geometry of the backgrounds up to local isometries.   As $\mathfrak{so}(N/4)$ is simple the universal cover of $G/H$ with $\mathfrak{Lie}(G)=\mathfrak{so}(N/4)$ is compact and homogeneous, see eg \cite{bohmkerr}. So  the internal space can be identified with the universal cover $\tilde G/\tilde H$ of $G/H$ for which $\tilde G$ can be chosen to be simply connected. }
\ref{table:nonlin} and \ref{table:nonlin7}.  Incidentally, this also means that if $N>16$ backgrounds were to exist, the R-symmetry group
of the dual CFT would have to act transitively on the internal space of the solution.


 A direct observation of the classification  of 6-dimensional homogeneous spaces $G/H$ in table \ref{table:nonlin} reveals that those that can occur
 as internal spaces of AdS$_4$ backgrounds with $N>16$ in 10 dimensions are
\bea
&&\mathrm{Spin}(7)/\mathrm{Spin}(6)~(N=28)~,~~~SU(4)/S(U(1)\times U(3))~(N=24)~,~~~
\cr
&&Sp(2)/U(2)~(N=20)~,~~~Sp(2)/(Sp(1)\times U(1)) ~(N=20)~,
\label{homo166}
\eea
where $N$ denotes the expected number of supersymmetries that can be preserved by the background and   we always take $G$ to be simply connected.  Observe  that there are no maximally supersymmetric AdS$_4$
solutions in 10-dimensional supergravities in agreement with the results of \cite{maxsusy}.
The proof of our result in IIB supergravity is based on a cohomological argument and does not use details of the 6-dimensional homogeneous spaces involved. However in (massive) IIA
supergravity, one has to consider details of the geometry of these coset spaces.
 Solutions with strictly $N=28$ and $N=20$ supersymmetries are ruled out
after a detailed analysis of the KSEs and dilaton field equation. In the standard IIA supergravity there is a solution with 24 supersymmetry and internal space locally isometric to the symmetric space
$SU(4)/S(U(1)\times U(3))=\mathbb{CP}^3$.  This solution   has already been  found in \cite{nillsonpope}.  The homogeneous space $Sp(2)/Sp(1)\times U(1)$, which is diffeomorphic to $\mathbb{CP}^3$, gives also a solution at a special
region  of the moduli space of parameters.  This solution  admits $24$ supersymmetries and  is locally isometric to that with internal space $SU(4)/S(U(1)\times U(3))$.


The classification  of 7-dimensional  homogeneous spaces $G/H$ in table  \ref{table:nonlin7}  reveals that those that can occur as internal spaces of $N>16$ AdS$_4$ backgrounds in   11 dimensions
 are
\bea
&&\mathrm{Spin}(8)/\mathrm{Spin}(7)~(N=32)~,~~\mathrm{Spin}(7)/G_2~(N=28)~,~~SU(4)/SU(3)~(N=24)~,~~~
\cr
&&Sp(2)/Sp(1)_{\text{max}}~(N=20)~, ~~Sp(2)/\Delta(Sp(1))~(N=20)~,~~
\cr
&&Sp(2)/Sp(1)~(N=20)~,
\label{homo167}
\eea
where $Sp(1)_{\mathrm{max}}$  and $\Delta(Sp(1))$ denote the maximal   and diagonal embeddings of $Sp(1)$ in $Sp(2)$, respectively, and $G$ is chosen to be simply connected.
It is known that there is a maximally supersymmetric solution AdS$_4\times S^7$ with internal
space $S^7=\mathrm{Spin}(8)/\mathrm{Spin}(7)$ \cite{fr, duffpopemax}.   After a detailed investigation of the
geometry of the above homogeneous spaces, the solutions of the KSEs and the warp factor field equation, one can also show that the rest of the coset spaces  do not give solutions with strictly 20, 24 and 28 supersymmetries.  However as the homogeneous spaces $Spin(7)/G_2$, $SU(4)/SU(3)$
 and $Sp(2)/Sp(1)$ are diffeomorphic to $S^7$, there is a region in the moduli space of their
  parameters which yields the maximally supersymmetric AdS$_4\times S^7$ solution.

The paper is organized as follows. In section 2, we  show that there are no IIB $N>16$ $AdS_4 \times _w M^6$ solutions. In section 3, we show that there is an up to an over scale unique solution  of IIA  supersgravity that preserves 24 supersymmetries.  In section 4, we demonstrate that all $N>16$ AdS$_4$ backgrounds of 11-dimensional supergravity are locally
isometric to the maximally supersymmetric $AdS_4\times S^7$ solution. In section 5 we state our conclusions. In appendix A, we explain our conventions, and in appendix B we summarize some aspects of the geometry
of homogeneous spaces that is used
throughout the paper.  In appendices C,D and E, we present some formulae for the homogeneous spaces that admit a transitive action of a group with Lie algebra  $\mathfrak{su}(k)$ or $\mathfrak{so}(5)=\mathfrak{sp}(2)$.

\newsection{\texorpdfstring{$N>16$ $AdS_4 \times _w M^6$}{AdS4xM6} solutions in IIB }

To investigate the IIB AdS$_4$ backgrounds, we shall use the approach and notation of \cite{iibads} where Bianchi identities,
  field equations and KSEs are first solved along the AdS$_4$ subspace of $AdS_4\times_w M^6$ and then
  the remaining independent conditions along the internal space $M^6$ are identified.   The bosonic fields of IIB supergravity are the metric,
  a complex 1-form field strength $P$, a complex 3-form field strength $G$ and a real  self-dual 5-form $F$. Imposing the symmetry of AdS$_4$ on the fields, one finds  that the metric and form field strengths are given by
\begin{align}
&ds^2 = 2 du (dr+rh) + A^2 (dz^2 +e^{2z/\ell}dx^2) + ds^2(M^6)~, \notag \\
&G = H, \quad P =\xi,  \quad F = A^2 e^{z/\ell} du \wedge (dr + rh) \wedge dz\wedge dx \wedge Y + *_6 Y~,
\end{align}
where the metric has been written as a near-horizon geometry \cite{adshor} with
\begin{align}
h = -\frac{2}{\ell} dz - 2 A^{-1} dA~.
\end{align}
The warp factor $A$ is a function on the internal manifold $M^6$, $H$ is the complex 3-form on $M^6$, $\xi$ is a complex 1-form on $M^6$ and $Y$ is a real 1-form on $M^6$. The $AdS_4$ coordinates are $(u, r, z, x)$ and we introduce the null-ortho-normal  frame
\begin{align}\label{frameiib}
\bbe^+ = du~, \quad \bbe^- = dr + r h~, \quad \bbe^z = A\, dz~, \quad \bbe^x = A e^{z/\ell}\, dx~, \quad \bbe^i= \bbe^i_I\, dy^I~,
\end{align}
where $ds^2(M^6)=\delta_{ij} \bbe^i \bbe^j$. All gamma matrices are taken with respect to this null ortho-normal  frame.

The  Bianchi identities along $M^6$ which are useful
in the  analysis that follows are
\begin{align} \label{bianchi}
d(A^4Y) &= 0, \quad dH = iQ \wedge H - \xi \wedge \overline{H}, \notag\\
\nabla^i Y_i &= -\frac{i}{288} \epsilon^{i_1 i_2 i_3 j_1 j_2 j_3} H_{i_1 i_2 i_3} \overline{H}_{j_1 j_2 j_3}~, \notag \\
dQ &= -i \xi \wedge \bar{\xi}~,
\end{align}
where $Q$ is the pull-back of the canonical connection of the upper-half plane on the spacetime with respect to the dilaton and axion scalars of IIB supergravity. Similarly, the field equations of the warp factor is
\begin{align}
A^{-1}\nabla^2 A = 4Y^2 + \frac{1}{48}  H_{i_1i_2i_3} \overline{ H}^{i_1i_2i_3} - \frac{3}{\ell^2} A^{-2} - 3 A^{-2} (dA)^2~,
\label{einsteinA}
\end{align}
and those of the scalar and 3-form fluxes are
\begin{align}\label{feqn}
\nabla^i \xi_i &= -3 \partial^i \log A \, \xi_i + 2i Q^i \xi_i - \frac{1}{24} H^2~, \notag \\
\nabla^i H_{ijk} &= -3 \partial^i \log A \, H_{ijk} + i Q^i H_{ijk} + \xi^i \overline{H}_{ijk}~.
\end{align}
The full set of Bianchi identities and field equations can be found in \cite{iibads}. Note in particular that \eqref{einsteinA} implies that if $A$ and the other fields are smooth, then $A$ is nowhere vanishing on $M^6$.

\subsection{The Killing spinors}

After solving the KSEs along AdS$_4$, the Killing spinors of the background can be written as
\begin{align}\label{killingspinorsiib}
\epsilon  = \,&\sigma_+ - \ell^{-1} x \Gamma_{xz} \tau_+ + e^{-\frac{z}{\ell}} \tau_+ + \sigma_- + e^{\frac{z}{\ell}} ( \tau_- - \ell^{-1} x \Gamma_{xz} \sigma_-) \notag\\
& - \ell^{-1} u A^{-1} \Gamma_{+z} \sigma_- - \ell^{-1} r A^{-1} e^{-\frac{z}{\ell}} \Gamma_{-z} ~, \tau_+~,
\end{align}
where we have used the light-cone projections
\begin{align}
\Gamma_\pm \sigma_\pm = 0~, \quad \Gamma_\pm \tau_\pm = 0~,
\end{align}
and $\sigma_\pm$ and $\tau_\pm$ are $Spin(9,1)$ Weyl spinors depending only on the coordinates of $M^6$.
The remaining independent KSEs are
\begin{align} \label{kseiib1}
\nabla^{(\pm)}_i \sigma_\pm = 0~, \quad \nabla^{(\pm)}_i \tau_\pm = 0~,
\end{align}
and
\begin{align}\label{algkseiib2}
\left(\frac{1}{24} \sH + \slashed{\xi} C* \right) \sigma_\pm = 0~, \quad \left(\frac{1}{24} \sH + \slashed{\xi} C* \right) \tau_\pm = 0~,
\end{align}
as well as
\begin{align} \label{algkseiib}
\Xi^{(\pm)} \sigma_\pm = 0~, \quad \left(\Xi^{(\pm)} \pm \frac{1}{\ell} \right) \tau_\pm = 0~,
\end{align}
where
\begin{align}
&\nabla^{(\pm)}_i = \nabla_i \pm \frac{1}{2} \partial_i \log A - \frac{i}{2}Q_i \mp \frac{i}{2} \sgY_i \Gamma_{xz} \pm \frac{i}{2} Y_i \Gamma_{xz} + \left(-\frac{1}{96} \sgH_i + \frac{3}{32} \sH_i \right) C *~, \\
&\Xi^{(\pm)} = \mp \frac{1}{2\ell} - \frac{1}{2} \Gamma_z \slashed{\partial}A \pm \frac{i}{2} A \Gamma_x \sY + \frac{1}{96} A \Gamma_z \sH C*~,
\end{align}
and $C*$ is the charge conjugation matrix followed by standard complex conjugation. For some explanation of the notation see appendix A. (\ref{kseiib1}) and (\ref{algkseiib2}) can be thought of as
   the naive restriction of gravitino and dilatino KSEs of IIB supergravity  on $M^6$, respectively.    (\ref{algkseiib}) are algebraic and arise
    as  integrability conditions of the integration of  IIB KSEs over the AdS$_4$ subspace of the background. We do not assume that the Killing spinors factorize as Killing spinors on $AdS_4$ and Killing spinors on the internal manifold. It has been observed in \cite{iibads} that if $\sigma_+$ is a Killing spinor,  then
\bea
\tau_+=\Gamma_{zx} \sigma_+~, ~~~\sigma_-=A\Gamma_{-z}\sigma_+~,~~~\tau_-=A \Gamma_{-x}\sigma_+~,
\label{ksrel}
\eea
are also Killing spinors. As a result  AdS$_4$ solutions  preserve $4k$ supersymmetries.


\subsection{The non-existence of \texorpdfstring{$N>16$ $AdS_4$}{N greater 16 AdS4} solutions in IIB}

\subsubsection{Conditions on spinor bilinears}

As it has already been mentioned,  the two assumptions we have made in the introduction are equivalent for all   IIB, (massive) IIA and 11-dimensional AdS$_4$ backgrounds that preserve $N>16$ supersymmetries.
   Hence in what follows, we shall focus only on the restrictions  on the geometry of the spacetime imposed  by the first assumption which requires that the solutions are smooth
   and the internal space is compact without boundary.

To begin our analysis, a consequence of the homogeneity theorem  \cite{homogen} for solutions which preserve $N>16$  supersymmetries is that the IIB
 scalars are constant which in turn implies that
\begin{align}
\xi=0~.
\end{align}
As $Q$ is the pull-back of the canonical connection of the upper half plane with respect to the scalars and these are constant,  $Q=0$ as well.

Setting  $\Lambda = \sigma_+ + \tau_+$ and after using the gravitino KSE \eqref{kseiib1}, we find
\begin{align}\label{hopf1}
\nabla_i \parallel \Lambda \parallel^2 = - \parallel \Lambda \parallel^2 A^{-1} \nabla_i A - i Y_i \langle \Lambda, \Gamma_{xz} \Lambda \rangle + \frac{1}{48} \text{Re} \langle \Lambda, \sgH_i C* \Lambda \rangle~.
\end{align}
Next, observe that the algebraic KSE \eqref{algkseiib} implies
\begin{align}
\frac{1}{48}\sH C * \Lambda = \left(A^{-1} \Gamma^j \nabla_j A + i \Gamma^j \Gamma_{xz} Y_j \right) \Lambda + \ell^{-1} A^{-1} \Gamma_z (\sigma_+ - \tau_+)~,
\end{align}
which, when substituted back into \eqref{hopf1}, yields
\begin{align}\label{hopf2}
\nabla_i \parallel \Lambda \parallel^2 = 2 \ell^{-1} A^{-1} \text{Re} \langle \tau_+, \Gamma_{iz} \sigma_+ \rangle~.
\end{align}
However, the gravitino KSE \eqref{kseiib1} also implies that
\begin{align}
\nabla^i \left( A \text{Re} \langle \tau_+, \Gamma_{iz} \sigma_+ \rangle \right) = 0~.
\label{iibortho1}
\end{align}
Thus, in conjunction with \eqref{hopf2}, we obtain
\begin{align}
\nabla^2 \parallel \Lambda \parallel^2 + 2 A^{-1} \nabla^i A \nabla_i \parallel \Lambda \parallel^2 = 0~.
\end{align}
The Hopf maximum principle then implies that $\parallel \Lambda \parallel^2$ is constant, so  \eqref{hopf1} and \eqref{hopf2} give the conditions
\begin{align}\label{hopf1.1}
- \parallel \Lambda \parallel^2 A^{-1} \nabla_i A - i Y_i \langle \Lambda, \Gamma_{xz} \Lambda \rangle + \frac{1}{48} \text{Re} \langle \Lambda, \sgH_i C* \Lambda \rangle = 0~,
\end{align}
and
\begin{align}\label{hopf2.1}
\text{Re} \langle \tau_+, \Gamma_{iz} \sigma_+ \rangle = 0~,
\end{align}
respectively.  The  above equation  can be equivalently  written as $\text{Re} \langle \sigma_+, \Gamma_{ix} \sigma_+ \rangle = 0$.

The spinors $\sigma_+$ and $\tau_+$ are linearly independent as it can be easily seen from \eqref{algkseiib}.  Moreover as a consequence of (\ref{hopf2.1}),  they are orthogonal
\begin{align}
\text{Re} \langle \tau_+, \sigma_+ \rangle = 0~.
\label{iibortho2}
\end{align}
 To see this take the real part of   $\langle \tau_+, \Xi^{(+)} \sigma_+ \rangle -\langle \sigma_+, (\Xi^{(+)}+ \ell^{-1}) \tau_+ \rangle = 0$.  The conditions (\ref{iibortho1}), (\ref{iibortho2}) as well as the constancy of $\parallel\Lambda \parallel$ can also be derived from the assumption that the
 isometries of the background decompose into those of AdS$_4$ and those of the internal manifold \cite{superalgebra}.

\subsubsection{The warp factor is constant and the 5-form flux vanishes}

  AdS$_4$ backgrounds preserving $4k$ supersymmetries admit $k$ linearly independent
Killing spinors $\sigma_+$.  For every pair of such spinors $\sigma^1_+$ and $\sigma^2_+$ define the bilinear

\begin{align}\label{KV}
W_i = A \, \text{Re} \langle \sigma^1_+, \Gamma_{iz} \sigma^2_+ \rangle~.
\end{align}
Then the gravitino KSE \eqref{kseiib1} implies that
\begin{align}
\nabla_{(i}W_{j)}=0~.
\end{align}
Therefore W is a Killing vector on $M^6$.

Next consider the algebraic KSE (\ref{algkseiib}) and take the real part of $\langle \sigma^1_+, \Xi^{(+)} \sigma^2_+ \rangle -\langle \sigma^2_+, \Xi^{(+)} \sigma^1_+ \rangle = 0$
to find that
\bea
W^i \, \nabla_i A =0~,
\label{iib44}
\eea
where we have used (\ref{hopf2.1}).

Similarly, taking the real part of the difference  $\langle \sigma^1_+, \Gamma_{zx}\Xi^{(+)} \sigma^2_+ \rangle -\langle \sigma^2_+, \Gamma_{zx}\Xi^{(+)} \sigma^1_+ \rangle = 0$
and after using the condition (\ref{iibortho2}), we find
\begin{align}
i_W Y=0~.
\label{iib44a}
\end{align}

The conditions (\ref{iib44}) and (\ref{iib44a}) are valid for  all IIB AdS$_4$ backgrounds.
However if the solution preserves more than 16 supersymmetries,  an argument similar to that used for the proof of the homogeneity theorem in  \cite{homogen} implies that
the Killing vectors $W$ span the tangent spaces of $M^6$ at each point.  As a result, we conclude that
\bea
dA=Y=0~.
\label{iibvan}
\eea
Therefore the warp factor $A$ is constant and the 5-form flux $F$ vanishes. So the background is a product $AdS_4\times M^6$, and as it has been explained in the introduction $M^6$ is one of
the homogeneous spaces in (\ref{homo166}).

\subsubsection{Proof of the main statement}
To begin, it has been shown in  \cite{Gran:2009cz} that all IIB AdS backgrounds that
preserve $N\geq 28$ supersymmetries are locally isometric to the maximally supersymmetric ones.
As there is not a maximally supersymmetric AdS$_4$ background in IIB, we conclude that there
does not exist a AdS$_4$ solution which preserves $N\geq 28$ supersymmetries.

To investigate the $N=20$ and $N=24$ cases, substitute (\ref{iibvan}) into the Bianchi identities and field equations to find that  $H$ is harmonic and
\begin{align}\label{Hsqu}
H^2 = 0~.
\end{align}
If $H$ were real, this condition  would have implied $H=0$ and in turn would have led to a contradiction.  This is  because  the field equation for the warp factor \eqref{einsteinA}  cannot be satisfied.  Thus  we can already exclude the existence of such backgrounds.

Otherwise for solutions to exist,  $M^6$ must be a compact, homogeneous,  6-dimensional Riemannian manifold whose de-Rham cohomology  $H^3(M^6)$ has at least two generators and which admits a transitive and effective action of a group with Lie algebra isomorphic to either  $\mathfrak{so}(6)$ or $\mathfrak{so}(5)$ for  $N=24$ and $N=20$, respectively \cite{superalgebra}.
The homogeneous spaces that admit a transitive and effective  action of $\mathfrak{so}(6)$ or $\mathfrak{so}(5)=\mathfrak{sp}(2)$ have already been listed in (\ref{homo166}) and none of them satisfies these
 cohomology criteria. All compact homogeneous 6-manifolds have been classified in \cite{niko6dim} and the complete list of the simply connected ones relevant here is given in table~\ref{table:nonlin}.  Therefore, we conclude that there do not exist AdS$_4$ backgrounds preserving $N>16$ supersymmetries in IIB supergravity\footnote{Note that the possibility of IIB $AdS_4\times Z\backslash G/H$  backgrounds preserving $N>16$ supersymmetry is  also excluded, where $Z$
 is a discrete subgroup of $G$,  as there are  no   IIB  $AdS_4\times  G/H$ local geometries that  preserve $N>16$ supersymmetries.}.

\begin{table}\renewcommand{\arraystretch}{1.3}
	\caption{6-dimensional compact, simply connected,  homogeneous spaces}
	\centering
	\begin{tabular}{c l}
		\hline
		& $M^6=G/H$  \\  
		\hline
		(1)& $\frac{\mathrm{Spin}(7)}{\mathrm{Spin}(6)}= S^6$, symmetric space\\
        (2) & $\frac{G_2}{SU(3)}$ diffeomorphic to $S^6$\\
		(3)&$\frac{SU(4)}{S(U(1)\times U(3))}=\mathbb{CP}^3$, symmetric space\\
		(4)& $\frac{Sp(2)}{U(2)}$, symmetric space \\
		(5) & $\frac{Sp(2)}{Sp(1)\times U(1)}$ diffeomorphic to $\mathbb{CP}^3$  \\
(6) & $\frac{SU(3)}{T_{max}}$ Wallach space\\
		(7) & $\frac{SU(2)\times SU(2)}{\Delta(SU(2))} \times \frac{SU(2)\times SU(2)}{\Delta(SU(2))} = S^3 \times S^3 $\\
		(8) & $SU(2) \times \frac{SU(2) \times SU(2)}{\Delta(SU(2))}$  diffeomorphic to $S^3\times S^3$\\
        (9) & $SU(2) \times SU(2) $ diffeomorphic to $S^3\times S^3$ \\
(10) & $\frac{SU(2)}{U(1)} \times \frac{SU(2)}{U(1)} \times \frac{SU(2)}{U(1)}= S^2 \times S^2 \times S^2$\\
		(11) & $\frac{SU(2)}{U(1)} \times\frac{\mathrm{Spin}(5)}{\mathrm{Spin}(4)} = S^2 \times S^4$  \\
(12) & $\frac{SU(2)}{U(1)} \times\frac{SU(3)}{S(U(1)\times U(2))} = S^2 \times \mathbb{CP}^2$  \\
 [1ex]
		\hline
	\end{tabular}
	\label{table:nonlin}
\end{table}

\newsection{\texorpdfstring{$N>16$ $AdS_4 \times_w M^6$}{AdS4xM6} solutions  in (massive) IIA }\label{IIA}

To begin, let us summarize the solution of Bianchi identities, field equations and KSEs for (massive) IIA  $AdS_4 \times_w M^6$ backgrounds as presented in \cite{iiaads} whose notation we follow.   The bosonic fields
of (massive) IIA supergravity are  the metric,  a 4-form field strength $G$, a 3-form field strength $H$, a 2-form field strength $F$, the dilaton $\Phi$ and  the mass parameter $S$ of massive IIA dressed with the dilaton. Imposing the symmetries of AdS$_4$ on the fields, one finds that
\bea
&&ds^2=2 \bbe^+ \bbe^- + (\bbe^z)^2 + (\bbe^x)^2 + ds^2(M^6)~,
\cr
&&G = X \bbe^+ \wedge \bbe^- \wedge \bbe^z \wedge \bbe^x + Y~,
\quad H =H, \quad F=F, \quad \Phi=\Phi, \quad S=S,
\eea
where $ds^2(M^6)= \delta_{ij} \bbe^i \bbe^j$ and the frame $(\bbe^+, \bbe^-, \bbe^x, \bbe^z, \bbe^i)$ is defined as in \eqref{frameiib}. Note that the fields $H$, $F$, $\Phi$ and $S$ do not have a component along AdS$_4$ and so we use the same symbol to denote them and their component along $M^6$.   The warp factor $A$, $S$ and $X$ are functions of $M^6$, whereas $Y$, $H$ and $F$ are 4-form, 3-form and 2-form fluxes on $M^6$, respectively.  The  conditions imposed on the fields by the Bianchi identities and field equations after solving along the AdS$_4$ subspace can  be found in \cite{iiaads}. Relevant to our analysis that follows are the Bianchi identities
\begin{align}\label{iiabianchi}
dH&=0, \quad dS=S d\Phi~, \quad dY= d\Phi \wedge Y+ H\wedge F~,\notag\\
 dF&= d\Phi \wedge F + SH~, \quad d(A^4 X)= A^4 d\Phi~,
\end{align}
and the field equations for the fluxes
\begin{align}\label{iiafieldeqs}
\nabla^2 \Phi &= -4A^{-1} \partial^i  A \, \partial_i \Phi + 2 (d\Phi)^2 + \frac{5}{4} S^2 + \frac{3}{8} F^2 -\frac{1}{12} H^2 + \frac{1}{96} Y^2 - \frac{1}{4} X^2~,\notag\\
\nabla^k H_{ijk} &= -4A^{-1} \partial^k  A \, H_{ijk} + 2 \partial^k\Phi H_{ijk} + S F_{ij} +\frac{1}{2} F^{k\ell} Y_{ijk\ell}~, \notag\\
\nabla^{j} F_{ij} &= -4A^{-1} \partial^j  A \, F_{ij} + \partial^j \Phi \, F_{ij} - \frac{1}{6} H^{jkl} Y_{ijkl}~, \notag\\
\nabla^{\ell} Y_{ijk\ell} &= -4A^{-1} \partial^\ell  A \, Y_{ijk\ell} + \partial^\ell \Phi Y_{ijk\ell}~,
\end{align}
along $M^6$. Moreover, we shall use  the field equation for the warp factor $A$ and the Einstein field
equation along $M^6$
\begin{align}\label{einstiia}
\nabla^2 \log A &= -\frac{3}{\ell^2 A^2} - 4 (d \log A)^2 + 2 \, \partial_i \log A \partial^i \Phi + \frac{1}{96} Y^2 + \frac{1}{4} X^2 + \frac{1}{4} S^2 + \frac{1}{8} F^2~, \notag\\
R^{(6)}_{ij} &= 4 \nabla_i \partial_j \log A + 4 \partial_i \log A \, \partial_j \log A + \frac{1}{12} Y^2_{ij} - \frac{1}{96} Y^2 \delta_{ij} + \frac{1}{4} X^2 \delta_{ij} - \frac{1}{4} S^2 \delta_{ij} \notag\\
&\quad +\frac{1}{4} H_{ij}^2 + \frac{1}{2} F^2_{ij} - \frac{1}{8} F^2 \delta_{ij} - 2 \nabla_i \nabla_j \Phi~,
\end{align}
where $\nabla$ and $R^{(6)}_{ij}$ denote the Levi-Civita connection and  the Ricci tensor of $M^6$, respectively.

\subsection{The Killing spinor equations}

The solution of  KSEs of (massive) IIA supergravity along the $AdS_4$ subspace   can again be written as \eqref{killingspinorsiib}, where now $\sigma_\pm$ and $\tau_\pm$ are $\mathfrak{spin}(9,1)$ Majorana  spinors  that satisfy the lightcone projections $\Gamma_\pm\sigma_\pm=\Gamma_\pm\tau_\pm=0$
 and   depend only on the coordinates of $M^6$. After the lightcone projections are imposed,   $\sigma_\pm$ and $\tau_\pm$ have 16 independent components. These satisfy   the gravitino KSEs
\begin{align}\label{iiakse}
\nabla^{(\pm)}_i \sigma_\pm = 0~, \quad \nabla^{(\pm)}_i \tau_\pm = 0~,
\end{align}
the dilatino KSEs

\begin{align}\label{iiadilat}
\mathcal{A}^{(\pm)} \sigma_{\pm} = 0~, \quad \mathcal{A}^{(\pm)} \tau_{\pm} = 0~,
\end{align}
and the algebraic KSEs

\begin{align}\label{iiaalgkse}
\Xi^{(\pm)}\sigma_{\pm}=0~, \quad (\Xi^{(\pm)} \pm \frac{1}{\ell})\tau_{\pm}=0~,
\end{align}
where

\begin{align}
\nabla_i^{(\pm)} &= \nabla_i \pm \frac{1}{2} \partial_i\log A + \frac{1}{8} \sH_i \Gamma_{11} + \frac{1}{8} S \Gamma_i + \frac{1}{16} \sF \Gamma_i \Gamma_{11} + \frac{1}{192} \sY \Gamma_i \mp \frac{1}{8} X \Gamma_{zxi}~, \notag \\
\mathcal{A}^{(\pm)} &= \slashed{\partial}\Phi + \frac{1}{12} \sH \Gamma_{11} + \frac{5}{4} S + \frac{3}{8} \sF \Gamma_{11} + \frac{1}{96} \sY \mp \frac{1}{4} X \Gamma_{zx}~, \notag \\
\Xi^{(\pm)} &= -\frac{1}{2\ell} + \frac{1}{2} \slashed{\partial}A \Gamma_z - \frac{1}{8} A S \Gamma_z - \frac{1}{16} A \sF \Gamma_z \Gamma_{11} - \frac{1}{192} A \sY \Gamma_z \mp \frac{1}{8} A X \Gamma_x~.
\end{align}
The first two equations arise from the naive restriction of the gravitino and dilatino KSEs of the theory on $\sigma_\pm$ and $\tau_\pm$, respectively,  while the last algebraic equation
is  an integrability condition that arises  from the integration of the IIA KSEs  on AdS$_4$.
As in the IIB case, the solutions of the above KSEs are related as in (\ref{ksrel}) and so such
backgrounds preserve $4k$ supersymmetries.

\subsection{\texorpdfstring{$AdS_4$}{AdS4} solutions with \texorpdfstring{$N>16$}{N greater 16} in IIA}

\subsubsection{Conditions on spinor bilinears}

The methodology to establish conditions on the Killing spinor bilinears which follow  from our assumption  that either the solutions are smooth and the internal space is compact without boundary
or that the even subalgebra of the Killing superalgebra decomposes as stated in the introduction is the same as that presented for  IIB. However, the formulae are somewhat different. Setting
$ \Lambda = \sigma_+ + \tau_+$ and upon
 using the gravitino KSE \eqref{iiakse}, one finds
\bea\label{iiahopf1}
\nabla_i \parallel \Lambda \parallel^2 = -\nabla_i \log A \parallel \Lambda \parallel^2 - \frac{1}{4} S \langle \Lambda, \Gamma_i \Lambda \rangle - \frac{1}{8} \langle \Lambda, \sgF_i \Gamma_{11} \Lambda \rangle - \frac{1}{96} \langle \Lambda, \sgY_i \Lambda \rangle~.
\eea
After multiplying the algebraic KSE \eqref{iiaalgkse} with  $\Gamma_{iz}$ on the other hand, one gets
\begin{align}
\frac{1}{2\ell} \langle \Lambda, \Gamma_{iz} (\sigma_+ - \tau_+) \rangle &= - \nabla_i A \parallel \Lambda \parallel^2 - \frac{A}{4} S \langle \Lambda, \Gamma_i \Lambda \rangle  - \frac{A}{8} \langle \Lambda, \sgF_i \Gamma_{11} \Lambda \rangle  \notag \\
 &\quad - \frac{A}{96} \langle \Lambda, \sgY_i \Lambda \rangle~.
\end{align}
Using this,  one can  rewrite \eqref{iiahopf1} as
\begin{align}\label{iiahopf2}
\nabla_i \parallel \Lambda \parallel^2 = \frac{2}{\ell A} \langle \tau_+, \Gamma_{iz} \sigma_+ \rangle~.
\end{align}
On the other hand the gravitino KSE  \eqref{iiakse} gives
\begin{align}
\nabla^i \left( A \langle \tau_+, \Gamma_{iz} \sigma_+ \rangle \right) = 0~.
\end{align}
Therefore taking the divergence of \eqref{iiahopf2}, one finds
\begin{align}
\nabla^2 \parallel \Lambda \parallel^2 + 2 \nabla^i \log A \, \nabla_i \parallel \Lambda \parallel^2 = 0~.
\end{align}
An application of the  Hopf maximum principle  gives  that $\parallel \Lambda \parallel^2$ is constant, which when inserted back into \eqref{iiahopf1} and \eqref{iiahopf2} yields

\begin{align}
 -\nabla_i \log A \parallel \Lambda \parallel^2 - \frac{1}{4} S \langle \Lambda, \Gamma_i \Lambda \rangle - \frac{1}{8} \langle \Lambda, \sgF_i \Gamma_{11} \Lambda \rangle - \frac{1}{96} \langle \Lambda, \sgY_i \Lambda \rangle = 0~,
\end{align}
and

\begin{align}\label{iiahopf3}
\langle \tau_+, \Gamma_{iz} \sigma_+ \rangle = 0~,
\end{align}
 respectively. The above condition can also be expressed as $\langle \sigma^1_+, \Gamma_{ix}\sigma^2_+ \rangle = 0$ for any two solutions $\sigma^1_+$ and $\sigma^2_+$ of the KSEs.

As in  IIB,  the algebraic KSE (\ref{iiaalgkse}) implies that  $\langle \tau_+, \Xi^{(+)} \sigma_+ \rangle -\langle \sigma_+, (\Xi^{(+)} + \ell^{-1}) \tau_+ \rangle = 0$.  This  together
with \eqref{iiahopf3}~ give  that $\langle \sigma_+, \tau_+ \rangle = 0$ and so the $\tau_+$ and $\sigma_+$ Killing spinors are orthogonal.

\subsubsection{The warp factor is constant}

To begin, for every pair of solutions $\sigma_+^1$ and $\sigma_+^2$ of the KSEs we define the 1-form bilinear

\begin{align}
W_i = A\, \mathrm{Im}\,\langle \sigma_+^1, \Gamma_{iz} \sigma_+^2 \rangle~.
\end{align}
Then the gravitino KSE \eqref{iiakse} implies that

\begin{align}
\nabla_{(i}W_{j)}=0~,
\end{align}
therefore $W$ is an Killing vector  on $M^6$.

Next  the difference  $\langle \sigma^1_+, \Xi^{(+)} \sigma^2_+ \rangle -\langle \sigma^2_+, \Xi^{(+)} \sigma^1_+ \rangle = 0$ implies that
\bea
W^i \, \nabla_i A =0~,
\label{wda}
\eea
where we have used (\ref{iiahopf3}).

So far we have not used that the solutions preserve $N>16$ supersymmetries. However if this is assumed, then (\ref{wda})  implies that the warp factor $A$ is constant.
This is a consequence of an adaptation of the homogeneity theorem on $M^6$. The homogeneity theorem also implies that $\Phi$ and $S$ are constant.  $X$ is also constant as a consequence of   the Bianchi identity \eqref{iiabianchi}.
Therefore we have established that if the backgrounds preserve $N>16$  supersymmetries, then
\bea
A=\mathrm{const}~,~~~\Phi=\mathrm{const}~,~~~S=\mathrm{const}~,~~~X=\mathrm{const}~.
\label{constsiia}
\eea
As the warp factor is constant, all backgrounds that preserve $N>16$  supersymmetries are
 products, $AdS_4\times M^6$. In addition  as it has been explained in the introduction,   $M^6$ is  a homogeneous space admitting a transitive and effective
 action of a group $G$ with Lie algebra $\mathfrak{so}(N/4)$. These homogeneous spaces have been listed in (\ref{homo166}).  In what follows, we shall explore all these 6-dimensional homogeneous spaces to search for IIA solutions
that preserve $N>16$ supersymmetries.

\subsection{\texorpdfstring{$N=28$}{N=28}}

There are no maximally supersymmetric AdS$_4$ backgrounds in (massive) IIA supergravity \cite{maxsusy}.  So the next case  to be investigated is that with 28 supersymmetries.
In such a case  $M^6$ admits a transitive and effective action of a group with Lie algebra $\mathfrak{so}(7)$.
Amongst the homogeneous spaces presented in (\ref{homo166}), the only one with this  property is  $\mathrm{Spin}(7)/\mathrm{Spin}(6)=S^6$.

As   $\mathrm{Spin}(7)/\mathrm{Spin}(6)=S^6$ is a symmetric space, all left-invariant forms are parallel with respect to the Levi-Civita connection and so represent classes in the de-Rham cohomology. As $H^2(S^6)=H^3(S^6)=H^4(S^6)=0$, one concludes
that $F=H=Y=0$.  Using this and (\ref{constsiia}),  the dilatino KSE \eqref{iiadilat} implies that
\begin{align}
\left(\frac{5}{4} S - \frac{1}{4} X \Gamma_{zx}\right) \sigma_+ =0~.
\end{align}
As it is the sum of two commuting terms one Hermitian and the other anti-Hermitian, the existence of solutions requires that both must vanish separately.  As a result  $S=X=0$.
Therefore all fluxes must vanish.  This in turn leads to a contradiction as the field equation of the warp factor (\ref{einstiia}) cannot admit any solutions. Thus there are no (massive) IIA $AdS_4$ backgrounds   preserving   28 supersymmetries.

\subsection{\texorpdfstring{$N=24$}{N=24}}\label{cp3}

The internal space of  AdS$_4$ backgrounds that preserve 24 supersymmetries admits a  transitive and effective action of a group with Lie algebra $\mathfrak{so}(6)=\mathfrak{su}(4)$.
The only space  in (\ref{homo166}) compatible with  such an action is  ${SU(4)}/{S(U(1)\times U(3))}=\mathbb{CP}^3$. Again this is a symmetric space and so
all invariant forms are parallel with respect to the Levi-Civita connection. In turn they represent classes in the de-Rham cohomology. As $H^{\mathrm{odd}}(\mathbb{CP}^3)=0$, this implies that $H=0$.

It is well-known that this homogeneous space is a K\"ahler manifold and the  left-invariant metric is given by the standard Fubini-Study metric on $\mathbb{CP}^3$. The even cohomology ring of $\mathbb{CP}^3$ is generated by the K\"ahler form $\omega$. As a result the 2- and 4-form fluxes can be written as
\begin{align}\label{iiafluxes}
F= \alpha \, \omega~,~~
Y=\frac{1}{2} \, \beta \,\omega\wedge \omega~,
\end{align}
for some real constants $\alpha$ and $\beta$ to be determined.

To determine $\alpha$ and $\beta$, let us first consider  the dilatino KSE \eqref{iiadilat} which after imposing (\ref{constsiia}) reads
\begin{align}
\left( \frac{5}{4} S + \frac{3}{8} \sF \Gamma_{11} + \frac{1}{96} \sY - \frac{1}{4} X \Gamma_{zx}\right) \sigma_ + = 0~.
\label{ksexxiia}
\end{align}
The Hermitian and anti-Hermitian terms in this equation commute and so they can be separately imposed. Notice that the only non-trivial commutator to check is $[\sF, \sY]$ which vanishes because $F$ is proportional to the K\"ahler form
while $Y$ is a (2,2)-form with respect to the associated complex structure. Thus we have
\begin{align}\label{2-formintegr}
\left(\frac{3}{8} \sF \Gamma_{11} -  \frac{1}{4} X \Gamma_{zx}\right) \, \sigma_+ = 0~,
\end{align}
and
\begin{align}\label{Ykse}
\left(\frac{5}{4} S + \frac{1}{96} \sY \right)\, \sigma_+ = 0~.
\end{align}
Inserting these into the algebraic KSE \eqref{iiaalgkse} simplifies to
\begin{align}
\left(3 S \Gamma_z -  X \Gamma_x\right) \, \sigma_+ = {3\over \ell A}\sigma_+~.
\label{kseiiaa2x}
\end{align}
The integrability condition of this yields
\begin{align}\label{Xsqu}
X^2+9 S^2 = \frac{9}{\ell^2 A^2}~.
\end{align}

Next let us focus on  (\ref{2-formintegr}) and \eqref{Ykse}.  Choosing without loss of generality $\Gamma_{11}= \Gamma_{+-} \Gamma_{zx} \Gamma_{123456}$,
(\ref{2-formintegr}) can be rewritten as
\begin{align}
\alpha( \Gamma^{3456}+\Gamma^{1256}+\Gamma^{1234})\sigma_+= -\frac{X}{3}  \sigma_+~,
\label{kseiia3x}
\end{align}
and similarly \eqref{Ykse} as

\begin{align}\label{Ykse2}
\beta(\Gamma^{1234}+ \Gamma^{1256} + \Gamma^{3456}) \sigma_+ = -5 S \sigma_+~,
\end{align}
where we have chosen an ortho-normal frame for which $\omega= {\bf e}^{12}+ {\bf e}^{34}+ {\bf e}^{56} $.

To solve (\ref{kseiia3x}) and \eqref{Ykse2}, we decompose $\sigma_+$ into eigenspaces of $J_1=\Gamma_{3456}$ and $J_2=\Gamma_{1256}$ and find that this leads to the relations
\begin{align}
\alpha = -\frac{1}{3} X, \quad \beta = -5S~,
\label{raa}
\end{align}
for the eigenspaces $|+,+\rangle$, $|+,-\rangle$, $|-,+\rangle$,  and
\begin{align}
\alpha = \frac{1}{9} X, \quad \beta = \frac{5}{3} S~,
\label{rbb}
\end{align}
for the eigenspace $|-,-\rangle$.

Before we proceed to investigate the  KSEs further, let us focus on the field equations for the fluxes and the warp factor. Observe that $\alpha\not=0$.  Indeed if $\alpha=0$,  then the KSEs would have implied that  $X=0$.  As $H=X=0$,   the dilaton  field equation in (\ref{iiafieldeqs})  implies  that
all fluxes vanish.  In such a case, the warp factor field equation in (\ref{einstiia}) cannot be satisfied.

Thus  $\alpha\not=0$.  Then the   field equation for the 3-form flux in (\ref{iiafieldeqs}) becomes   $\alpha (S+ 4 \beta)=0$ and so this implies that
 $\beta=-1/4\, S$.  This contradicts the  results from  KSEs in (\ref{raa}) and (\ref{rbb}) above unless  $\beta=S=0$.  Setting $S=Y=0$ in the dilaton field equation in (\ref{iiafieldeqs}),
it is easy to see that it is satisfied if and only if   $\alpha=-1/3 X$ and so $\sigma_+$ lies in the eigenspaces $|+,+\rangle$, $|+,-\rangle$ and  $|-,+\rangle$. As $S=0$,  (\ref{Xsqu}) implies that $X= \pm 3 \ell^{-1} A^{-1}$ and so  $\alpha=\mp \ell^{-1} A^{-1}$. The algebraic KSE (\ref{kseiiaa2x}) now reads $\Gamma_x \sigma_+ = \mp \sigma_+$.  As  $\alpha=-1/3 X$,  the common eigenspace  of $\Gamma_x$,   $\Gamma_{3456}$ and $\Gamma_{1256}$ on $\sigma_+$ spinors
has dimension 6. Thus the number of supersymmetries that the background
\bea
&&ds^2=2 du (dr-2 \ell^{-1} r dz) + A^2 (dz^2 +e^{2z/\ell}dx^2) + ds^2(\mathbb{CP}^3)~,
\cr
&&G = \pm 3 \ell^{-1} A e^{z/\ell} du \wedge dr \wedge dz \wedge dx~,
\quad H = S=0~,
\cr
&& F=\mp \ell^{-1} A^{-1} \omega, \quad \Phi=\mathrm{const}~,
\label{24soliia}
\eea
with $R^{(6)}_{ij} \delta^{ij}=24 \ell^{-2}A^{-2}$,  can  preserve is  24.

To establish that (\ref{24soliia}) preserves 24 supersymmetries, it remains to investigate the gravitino KSE \eqref{iiakse}. As  $\mathbb{CP}^3$ is simply connected it is sufficient to investigate the integrability condition
\begin{align}
\left( \frac{1}{4} R_{ijmn} \Gamma^{mn} - \frac{1}{8} F_{im} \, F_{jn} \Gamma^{mn} -\frac{1}{12} X\, F_{ij} \Gamma_{zx} \Gamma_{11} -\frac{1}{72} X^2 \Gamma_{ij} \right) \, \sigma_+ = 0~,
\label{intcp3}
\end{align}
of the gravitino KSE. The Riemann tensor of ${SU(4)}/{S(U(1)\times U(3))}$  is
\bea
R_{ij,kl}=\frac{1}{4 \ell^2 A^2} (\delta_{ik} \delta_{jl}-\delta_{il} \delta_{jk})+ \frac{3}{4 \ell^2 A^2} (\omega_{ij} \omega_{kl}- \omega_{i[j} \omega_{kl]})~.
\label{cp3curv}
\eea
Then a  substitution of this and the rest of the fluxes into the integrability condition reveals that it is satisfied without further conditions. In a similar manner, one can check that the Einstein equation along $M^6$ is also satisfied. This
is the IIA $N=24$  solution of  \cite{nillsonpope, dufflupope}.

\subsection{\texorpdfstring{$N=20$}{N=20}}\label{N=20}

The internal space of AdS$_4$ backgrounds that preserve 20 supersymmetries admits an effective and  transitive action of a group which has Lie algebra $\mathfrak{so}(5)=\mathfrak{sp}(2)$.
An inspection of the homogeneous spaces in table~\ref{table:nonlin} reveals that there are two candidate internal spaces namely the symmetric space ${Sp(2)}/{U(2)}$ and the homogeneous space ${Sp(2)}/{Sp(1)\times U(1)}$. The symmetric space is the space of complex structures on $\mathbb{H}^2$ which are compatible with the quaternionic inner product while the homogeneous space
is identified with the coset space of the sphere $\bar x x+ \bar y y=1$, $x,y\in \mathbb{H}$, with respect to the action $(x, y)\rightarrow (a x, ay)$, $a\in U(1)$.
The latter is diffeomorphic to $\mathbb{CP}^3$.

 \subsubsection{\texorpdfstring{${Sp(2)}/{U(2)}$}{Sp(2) over U(2)}}\label{Sp2overU2}

The geometry and algebraic properties of this symmetric space are described in appendix E. The most general left-invariant metric is
\bea
ds^2=a\,\delta_{rs} \delta_{ab} \bbl^{ra} \bbl^{sb}=\delta_{rs} \delta_{ab} {\bf e}^{ra} {\bf e}^{sb}~,
\label{sp2u2metr}
\eea
where $a>0$ is a constant and $\bbl^{ra}$, and ${\bf e}^{ra}=\sqrt{a}\, \bbl^{ra}$ are the left-invariant and ortho-normal frames, respectively,
and where $r,s=1,2,3$ and $a,b=4,5$.
The invariant forms are generated by the 2-form
\bea
\omega=\frac12 \delta_{rs} \epsilon_{ab}\, {\bf e}^{ra}\wedge {\bf e}^{sb}~.
\label{kahlersp2}
\eea
  ${Sp(2)}/{U(2)}$ is  a K\"ahler manifold with respect to the pair $(ds^2, \omega)$.

To continue we choose the metric on the internal manifold as (\ref{sp2u2metr}) and the fluxes as in the ${SU(4)}/{S(U(1)\times U(3))}$ case,  i.e.
 \bea
 F= \alpha \, \omega~,~~~Y= {\frac{1}{2}}\beta \, \omega\wedge \omega~,
 \eea
but now $\omega$ is given in (\ref{kahlersp2}),  where $\alpha$ and $\beta$ are constants. Since there are no invariant 3-forms on ${Sp(2)}/{U(2)}$, this implies $H = 0$. Performing a similar analysis to that in section \ref{cp3}, we find that  $\beta=S=0$, $\alpha={\mp} \ell^{-1} A^{-1}$ and $X= \pm 3 \ell^{-1} A^{-1}$,
 and $\sigma_+$ to satisfy  the same Clifford algebra projections as in e.g. (\ref{kseiia3x}). This requires an appropriate re-labeling of the indices of the ortho-normal frame $\bbe^{ra}$ so that the left-invariant tensors take the same canonical form as those of  $SU(4)/S(U(1)\times U(3))$ expressed in terms of the  ortho-normal frame $\bbe^i$. As a result, there are 24 spinors that solve the  KSEs so far.

 It remains to investigate the solutions
 of the  gravitino KSE \eqref{iiakse}. As in the  $SU(4)/S(U(1)\times U(3))$ case in section \ref{cp3}, we shall investigate the integrability condition instead. This   is again given as in (\ref{intcp3}).
 The curvature of the metric of this symmetric space is presented  in (\ref{curvsp2u2}).
 Using this the integrability condition (\ref{intcp3}) is written as
 \bea
&&\big[ \frac{1}{16 a} (\delta^{cd} \Gamma_{rcsd}- \delta^{cd} \Gamma_{scrd})\delta_{ab}+ \frac{1}{16 a} \delta^{tu}( \Gamma_{taub}-\Gamma_{tbua}) \delta_{rs}-
\cr &&
 \frac18 \ell^{-2} A^{-2} (\delta^{cd} \Gamma_{rcsd}\delta_{ab}-  \Gamma_{sb ra})+\frac14 \ell^{-2} A^{-2} \delta_{rs} \epsilon_{ab} \Gamma_{zx} \Gamma_{11}
 \cr &&
 -\frac18 \ell^{-2} A^{-2}
 \Gamma_{rasb}\big]\sigma_+=0~.
 \label{innxxx}
 \eea
 Contracting with $\delta_{ab}$, one finds that there are solutions which preserve more than 8 supersymmetries provided $a=\ell^{2} A^{2}$.
 Then taking the trace of (\ref{innxxx}) with $\epsilon_{ab} \delta_{rs}$, we find that
 \bea
 \frac12\slashed{\omega}\sigma_+=-{12}\Gamma_{zx} \Gamma_{11} \sigma_+~,
 \eea
 which is in contradiction to the condition (\ref{2-formintegr})  arising from the dilatino KSE.   The symmetric space ${Sp(2)}/{U(2)}$ does not yield\footnote{${Sp(2)}/{U(2)}$ can also be excluded
 as a solution because it is not a spin manifold \cite{klausthesis}.}  AdS$_4$ solutions that preserve 20 supersymmetries.

\subsubsection{\texorpdfstring{${Sp(2)}/({Sp(1)\times U(1)})$}{Sp(2) over Sp(1)xU(1)}}\label{Sp2_over_Sp1xU1}

The ${Sp(2)}/({Sp(1)\times U(1)})$ homogeneous space is  described in appendix E. Introducing the  left-invariant frame $\bbl^A m_A= \bbl^a W_a+ \bbl^{\underline r} T^{(+)}_{\underline r}$, the most general left-invariant metric is
\bea
ds^2= a\, \delta_{ab} \bbl^a \bbl^b+ b \,\delta_{{\underline r}{\underline s}} \bbl^{\underline r}\bbl^{\underline s} = \delta_{ab} \bbe^a \bbe^{b} + \delta_{{\underline r}{\underline s}} \bbe^{{\underline r}} \bbe^{{\underline s}},
\eea
where we have introduced the ortho-normal frame $\bbe^a=\sqrt{a}\,\bbl^a$ , $\bbe^{{\underline r}}=\sqrt{b}\,  \bbl^{\underline r}$, and where ${\underline r}=1,2$ and $a,b=1,\dots, 4$. The invariant forms are generated by
\bea
I^{(+)}_3={1\over2} (I^{(+)}_3)_{ab} \bbe^a\wedge \bbe^b ~,~~~\tilde{\omega} = \frac12  \epsilon_{{\underline r}{\underline s}} \bbe^{\underline r}\wedge \bbe^{\underline s}~,~~~ \bbe^{\underline r}\wedge I^{(+)}_{\underline r}~,~~~
\eea
and their duals, where
\bea
I^{(+)}_{\underline r}={1\over2} (I^{(+)}_{\underline r})_{ab} \bbe^a\wedge \bbe^b~.
\eea
The matrices $\big((I^{(\pm)}_r)_{ab}\big)$ are a basis in the space of self-dual and anti-self dual 2-forms in $\bR^4$ and are defined in (\ref{Imatrpm}).  Imposing the  Bianchi identities \eqref{iiabianchi}, one finds   the relation
\begin{align}\label{sp2bianchi}
\frac{\alpha}{\sqrt{b}} - \frac{\beta\sqrt{b}}{2a} = S \, h~,
\end{align}
and that the fluxes can be written as
\begin{align}\label{iiasp2fluxes}
F &= \alpha I^{(+)}_3 + \beta \, \tilde{\omega}~, \quad H = h \, \epsilon_{{\underline r}{\underline s}} \, \bbe^{\underline{r}} \wedge I^{(+)}_{\underline{s}}~,\notag \\
Y&=\gamma \, \tilde{\omega}\wedge I^{(+)}_{3} +  \frac{1}{2} \, \delta \, I^{(+)}_{3} \wedge I^{(+)}_{3}~,
\end{align}
where $\alpha, \beta, h, \gamma$ and $\delta$ are constants.

The   dilatino KSE \eqref{iiadilat} is the sum of hermitian and anti-hermitian Clifford algebra elements  which commute and thus lead to the two independent conditions
\begin{align} \label{splitkse}
\left(\frac{3}{8} \sF \Gamma_{11} -  \frac{1}{4} X \Gamma_{zx}\right) \, \sigma_+ &= 0~, \notag \\
\left(\frac{5}{4}\, S + \frac{1}{12} \sH \Gamma_{11} + \frac{1}{96} \sY   \right) \, \sigma_+ &=0~.
\end{align}
Using this to simplify the algebraic KSE \eqref{iiaalgkse}, one finds

\begin{align}
\left( \frac{1}{12} \sH \, \Gamma_{11} \Gamma_z + S \Gamma_z - \frac{X}{3} \Gamma_{x}\right) \sigma_+ = \frac{1}{\ell A} \, \sigma_+~.
\label{kkkkk}
\end{align}
If we then insert the fluxes \eqref{iiasp2fluxes} into the above KSEs and set $J_1=\Gamma^{24\underline{1}}\Gamma_{11}$, $J_2=\Gamma^{13\underline{1}}\Gamma_{11}$ and $J_3=\Gamma^{23\underline{2}}\Gamma_{11}$, we obtain

\begin{align}
\left(\alpha (J_2 J_3 - J_1 J_3) + \beta J_1 J_2\right) \, \sigma_+ + \frac{X}{3} \sigma_+ &= 0~, \notag\\
\left( 5S + 2 h (J_1 -J_2-J_3 + J_1 J_2 J_3) + \gamma \, (J_2J_3 - J_1 J_3) + \delta \, J_1J_2   \right) \sigma_+ &=0~,\notag\\
\left( \frac{1}{2} h (J_1 -J_2-J_3 + J_1 J_2 J_3) \Gamma_z + S \Gamma_z - \frac{X}{3} \Gamma_{x} \right) \sigma_+ - \frac{1}{\ell A} \, \sigma_+ &= 0~.
\label{j1j2j3377}
\end{align}
As $J_1, J_2, J_3$  are commuting Hermitian Clifford algebra operators with eigenvalues
$\pm1$, the KSE (\ref{kkkkk}) can be decomposed along the common eigenspaces as described  in  table \ref{table2xx}.

\begin{table}[h]
\begin{center}
\vskip 0.3cm
 \caption{Decomposition of (\ref{j1j2j3377}) KSE into eigenspaces}
 \vskip 0.3cm
 \begin{tabular}{|c|c|c|}
		\hline
		&$|J_1,J_2,J_3\rangle$&  relations for the fluxes\\
		\hline
		(1)&$|+,+,+\rangle$, $|-,-,-\rangle$& $\beta=-\frac{X}{3}$, $5S + \delta = 0$ \\
		& $|+,+,-\rangle$, $|-,-,+\rangle$ &  $(S \, \Gamma_z - \frac{X}{3} \Gamma_{x} ) |\cdot\rangle = \frac{1}{\ell A} | \cdot \rangle$ \\
		\hline
		(2)&$|+,-,+\rangle$, $|-,+,-\rangle$& $2\alpha + \beta=\frac{X}{3}$, $5S - 2 \gamma - \delta = 0$ \\
		& & $(S \, \Gamma_z - \frac{X}{3} \Gamma_{x} ) |\cdot\rangle = \frac{1}{\ell A} | \cdot \rangle$ \\
		\hline
		(3)&$|+,-,-\rangle$	& $2 \alpha - \beta=-\frac{X}{3}$, $5S + 8h +2\gamma-\delta=0$ \\
		& & $((S+2h) \, \Gamma_z - \frac{X}{3} \Gamma_{x} ) |\cdot\rangle = \frac{1}{\ell A} | \cdot \rangle$ \\
		\hline
		(4)&$|-,+,+\rangle$	& $2 \alpha - \beta= -\frac{X}{3}$, $5S-8h + 2\gamma -\delta=0$ \\
		& & $((S-2h) \, \Gamma_z - \frac{X}{3} \Gamma_{x} ) |\cdot\rangle = \frac{1}{\ell A} | \cdot \rangle$ \\
		\hline
	\end{tabular}
 \vskip 0.2cm
  \label{table2xx}
 \end{center}
\end{table}

\vskip 0.5cm

From the results of table \ref{table2xx}, there are two possibilities to choose five $\sigma_+$  Killing spinors, namely those in eigenspaces (1) and (3) and those in eigenspaces (1) and (4). For both of these choices,  the Bianchi identity \eqref{sp2bianchi} and the dilaton field equation give
\begin{align}
\alpha=\beta=-\frac{X}{3}, \quad X= \pm \frac{3}{\ell A}, \quad b=2a, \quad S=h=\gamma=\delta=0~.
\label{4x4c}
\end{align}
In either case notice that  these conditions imply the existence of  six $\sigma_+$  Killing spinors as the conditions
required for both $|+,-,-\rangle$ and $|-,+,+\rangle$ to be solutions are satisfied.  So potentially this background can preserve $N=24$ supersymmetries.
To summarize, the independent conditions
 on the Killing spinors arising from those  in  (\ref{splitkse}) and those in  table
 \ref{table2xx} are
\begin{align}\label{sp2ksproject}
\frac{1}{2} \left( \slashed{I^{(+)}_3} + \tilde{\slashed{\omega}}\right) \sigma_+ = \sigma_+~, \quad \Gamma_x \sigma_+ = -\frac{3}{\ell A X} \sigma_x~.
\end{align}
These are  the same conditions as those found in section \ref{cp3} for $M^6=\mathbb{CP}^3$.

It remains  to investigate the gravitino KSE \eqref{iiakse} or equivalently, as  ${Sp(2)}/(Sp(1)\times U(1))$ is simply connected, the corresponding integrability condition given again in  \eqref{intcp3}.  The curvature of the metric is given in (\ref{curvsp2sp1u1}). Moreover the Einstein equation \eqref{einstiia} gives $a=\ell^2 A^2/2$. Using these and substituting  the  conditions  (\ref{4x4c}) into the integrability condition, one can show that this is automatically satisfied provided that  \eqref{sp2ksproject} holds. As a result, there are no AdS$_4$ backgrounds with
internal space ${Sp(2)}/({Sp(1)\times U(1)}$ which preserve {\sl strictly } 20 supersymmetries.  However as shown above,  there is a solution which preserves 24 supersymmetries for $b=2a$.  This   is locally isometric to the  $AdS_4\times \mathbb{CP}^3$ solution found in section \ref{cp3}.  Note that there are no $N>24$ solutions  as it can be seen by a direct
computation or by observing that $\mathbb{CP}^3$ does not admit an effective and transitive action by the  $\mathfrak{so}(N/4)$ subalgebra of the Killing
superalgebra of such backgrounds. However there are  $AdS_4\times {Sp(2)}/({Sp(1)\times U(1))}$ solutions which preserve 4 supersymmetries \cite{lust}.

\newsection{\texorpdfstring{$N>16$ $AdS_4\times_w M^7$}{AdS4} solutions in 11 dimensions}\label{11d}

\subsection{\texorpdfstring{$AdS_4$}{AdS4xM7} solutions in \texorpdfstring{$D=11$}{D=11}}

Let us first summarize some of the properties  of  $AdS_4\times_w M^7$ backgrounds  in 11-dimensional supergravity   as  described in  \cite{mads} that  we shall use later. The bosonic fields are given as
\begin{align}
ds^2&= 2 \bbe^+ \bbe^- + (\bbe^z)^2 + (\bbe^x)^2 + ds^2(M^7) ~, \notag\\
 F&= X\, \bbe^+\wedge \bbe^- \wedge \bbe^z \wedge \bbe^x + Y~,
\end{align}
where the null ortho-normal  frame $(\bbe^+, \bbe^-, \bbe^z, \bbe^x, \bbe^i)$ is as  in (\ref{frameiib}), but now $i,j=1,\dots, 7$, and the metric on the internal space $M^7$ is $ ds^2(M^7)= \delta_{ij} \bbe^i \bbe^j$.  $X$ and $Y$ are a function and 4-form on $M^7$, respectively.

The Bianchi identities of the 11-dimensional supergravity evaluated on the $AdS_4\times_w M^7$ background yield
\begin{align}\label{mbianchi}
dY = 0, \quad d(A^4 X)=0~.
\end{align}
Similarly, the field equations give
\begin{align}\label{Xfieldeqn}
\nabla^k Y_{ki_1i_2i_3} + 4 \nabla^k A \, Y_{ki_1i_2i_3} = - \frac{1}{24} X \epsilon_{i_1i_2i_3}{}^{k_1k_2k_3k_4} Y_{k_1k_2k_3k_4}~,
\end{align}
\begin{align}\label{m-einst-ads}
\nabla^k \partial_k \log A = - \frac{3}{\ell^2 A^2} - 4 \partial_k \log A \, \partial^k \log A + \frac{1}{3} X^2 + \frac{1}{144} Y^2~,
\end{align}
and
\begin{align}\label{m-einst-trans}
R^{(7)}_{ij} - 4 \nabla_i \partial_j \log A - 4 \partial_i \log A \partial_j \log A = \frac{1}{12} Y^2_{ij} + \delta_{ij} \left(\frac{1}{6} X^2 - \frac{1}{144} Y^2 \right)~,
\end{align}
where $\nabla$ is the Levi-Civita connection on $M^7$.

\subsection{The Killing spinors}

The solution of the KSEs of $D=11$ supergravity  along the AdS$_4$ subspace of $AdS_4\times_w M^7$  given in \cite{mads} can be expressed as in (\ref{killingspinorsiib}) but now
$\sigma_\pm$ and $\tau_\pm$ are $\mathfrak{spin}(10,1)$ Majorana  spinors  that depend on the coordinates of $M^7$.  Again they satisfy the lightcone projections $\Gamma_\pm\sigma_\pm=\Gamma_\pm\tau_\pm=0$.
The remaining independent KSEs are
\begin{align} \label{mkse}
\nabla^{(\pm)}_i \sigma_\pm = 0~, \quad \nabla^{(\pm)}_i \tau_\pm = 0~,
\end{align}
and

\begin{align}\label{malgkse}
\Xi^{(\pm)}\sigma_{\pm}=0~, \quad (\Xi^{(\pm)} \pm \frac{1}{\ell})\tau_{\pm}=0~,
\end{align}
where
\begin{align}
\nabla_i^{(\pm)} &= \nabla_i \pm \frac{1}{2} \partial_i \log A - \frac{1}{288} \sgY_i + \frac{1}{36} \sY_i \pm \frac{1}{12} X \Gamma_{izx}~, \\
\Xi^{(\pm)} &= \mp \frac{1}{2\ell} - \frac{1}{2} \Gamma_z \slashed{\partial} A + \frac{1}{288} A \Gamma_z \sY \pm \frac{1}{6} A X \Gamma_x~.
\end{align}
The former KSE is the restriction of the gravitino KSE on $\sigma_\pm$ and $\tau_\pm$ while the latter arises as an integrability condition as a result of integrating the gravitino KSE
of 11-dimensional supergravity
over the AdS$_4$ subspace of $AdS_4\times_w M^7$.

\subsection{\texorpdfstring{$AdS_4$}{AdS4} solutions with \texorpdfstring{$N>16$}{N greater 16} in 11 dimensions}

\subsubsection{Conditions on spinor bilinears}

The conditions that arise from the assumption that $M^7$ be compact without boundary and the solutions be smooth
are similar to those presented in the (massive) IIA case. In particular, one finds
\bea
\parallel \sigma_+\parallel=\mathrm{const}~,~~~\langle \tau_+, \Gamma_{iz} \sigma_+ \rangle = 0~,~~~  \langle \sigma_+, \tau_+ \rangle = 0~.
\eea
The proof follows the same steps as in the (massive) IIA case and so we shall not repeat it here.

\subsubsection{The warp factor is constant}

Using arguments  similar to those presented in the (massive) IIA case, one finds that
 $W_i = A\, \mathrm{Im} \,\langle \sigma_+^1, \Gamma_{iz} \sigma_+^2 \rangle$ are Killing vectors on $M^7$ for any pair of Killing spinors $\sigma_+^1$ and $\sigma_+^2$ and that $i_W dA=0$.

Next, let us suppose  that  the backgrounds preserve $N>16$  supersymmetries. In such a case a similar argument to that presented for the proof
  of the homogeneity conjecture implies that the $W$ vector fields span the tangent space of $M^7$ at every point and so $A$ is constant.
  From the Bianchi identity \eqref{mbianchi} it then follows that $X$ is constant as well.
  Thus we have established that
\begin{align}\label{AXconst}
A=\mathrm{const}~,~~~X=\mathrm{const}~.
\end{align}
As a result, the space time is a product $AdS_4\times M^7$, where $M^7$ is a homogeneous space.
Further progress requires the investigation of individual homogeneous spaces of dimension 7 which have been classified in  \cite{niko7dim, bohmkerr} and they
are presented in table~\ref{table:nonlin7}.  Requiring in addition that the homogeneous spaces which can occur as internal spaces of $N>16$ AdS$_4$ backgrounds  must admit an effective and transitive action of a group that has Lie algebra $\mathfrak{so}(N/4)$, one arrives at the homogeneous
spaces presented in (\ref{homo167}).  In what follows, we shall investigate in detail the geometry of these homogeneous spaces to search for $N>16$ AdS$_4$ backgrounds
in 11-dimensional supergravity.

\begin{table}\renewcommand{\arraystretch}{1.3}
	\caption{7-dimensional compact, simply connected,  homogeneous spaces}
	\centering
	\begin{tabular}{c l}
		\hline
		& $M^7=G/H$  \\  
		\hline
		(1)& $\frac{\mathrm{Spin}(8)}{\mathrm{Spin}(7)}= S^7$, symmetric space\\
		(2)&$\frac{\mathrm{Spin}(7)}{G_2}=S^7$ \\
		(3)& $\frac{SU(4)}{SU(3)}$ diffeomorphic to $S^7$ \\
        (4) & $\frac{Sp(2)}{Sp(1)}$ diffeomorphic to $S^7$ \\
		(5) & $\frac{Sp(2)}{Sp(1)_{max}}$, Berger space \\
		(6) & $ \frac{Sp(2)}{\Delta(Sp(1))}=V_2(\bR^5)$ \\
        (7) & $\frac{SU(3)}{\Delta_{k,l}(U(1))}=W^{k,l}$~~ $k, l$ coprime, Aloff-Wallach space\\
		(8)&$\frac{SU(2) \times SU(3) }{\Delta_{k,l}(U(1))\cdot (1\times SU(2))}=N^{k,l}$ ~$k,l$ coprime\\
		(9) & $\frac{SU(2)^3}{\Delta_{p,q,r}(U(1)^2)}=Q^{p,q,r}$ $p, q, r$ coprime\\
(10)&$M^4\times M^3$,~~$M^4=\frac{\mathrm{Spin}(5)}{\mathrm{Spin}(4)}, ~\frac{ SU(3)}{S(U(1)\times U(2))}, ~\frac{SU(2)}{U(1)}\times \frac{SU(2)}{U(1)}$\\
&~~~~~~~~~~~~~~~~$M^3= SU(2)~,~\frac{SU(2)\times SU(2)}{\Delta(SU(2))}$\\
(11)&$M^5\times \frac{SU(2)}{U(1)}$,~~$M^5=\frac{\mathrm{Spin}(6)}{\mathrm{Spin}(5)}, ~\frac{ SU(3)}{SU(2)}, ~\frac{SU(2)\times SU(2)}{\Delta_{k,l}(U(1))},~ \frac{ SU(3)}{SO(3)} $\\
[1ex]
		\hline
	\end{tabular}
	\label{table:nonlin7}
\end{table}

\subsection{\texorpdfstring{$N=28,~{Spin(7)}/{G_2}$}{N=28, Spin(7) over G2}}

The maximally supersymmetric solutions have been classified before \cite{maxsusy} where it has been shown that all are locally isometric to $AdS_4\times S^7$ with $S^7=\mathrm{Spin}(8)/\mathrm{Spin}(7)$. The only  solution that may preserve $N=28$ supersymmetries is associated with the homogeneous space ${Spin(7)}/{G_2}$, see (\ref{homo167}). The Lie algebra $\mathfrak{spin}(7)= \mathfrak{so}(7)$ is again spanned by matrices $M_{ij}$ as in \eqref{gen-sona} satisfying the commutation relations \eqref{son-commuta} where now $i, j=1,2,...,7$. Let us denote the generators of  $\mathfrak{g}_2$ subalgebra of $\mathfrak{spin}(7)$  and those of the module $\mathfrak{m}$, $\mathfrak{spin}(7)=\mathfrak{g}_2\oplus \mathfrak{m}$,   with $G$ and $A$, respectively. These are defined as

\begin{align}
G_{ij}= M_{ij} + \frac{1}{4} \ast_{{}_7}\!\varphi_{ij}{}{}^{kl}\, M_{kl}~, \quad A_i = \varphi_{i}{}^{jk} M_{jk}~,
\end{align}
where  $\varphi$ is the fundamental $G_2$ 3-form,  $\ast_{{}_7}\varphi$ is its dual and $\ast_{{}_7}$ is the duality operation along the 7-dimensional internal space.   The non-vanishing components of $\varphi$ and $\ast_{{}_7}\varphi$ can be chosen as
\begin{align}
\varphi_{123}&=\varphi_{147}=\varphi_{165}=\varphi_{246}=\varphi_{257}=\varphi_{354}=\varphi_{367}=1~, \notag\\
\ast_{{}_7}\varphi_{1276}&=\ast_{{}_7}\varphi_{1245}=\ast_{{}_7}\varphi_{1346}=\ast_{{}_7}\varphi_{1357}=\ast_{{}_7}\varphi_{2374}=\ast_{{}_7}\varphi_{2356}=\ast_{{}_7}\varphi_{4567}=1~,
\label{g2phi}
\end{align}
and we have raised the indices above using the flat metric.
We have used the conventions for  $\varphi$ and $\ast_{{}_7}\varphi$  of \cite{Gran:2016tqd}, where also several useful identities satisfied by  $\varphi$ and $\ast_{{}_7}\varphi$  are presented.
In particular observe that $\varphi_{i}{}^{jk} G_{jk} = 0$.
The $\mathfrak{spin}(7)$ generators can be written as
\begin{align}
M_{ij} = \frac{2}{3} \, G_{ij} + \frac{1}{6} \varphi_{ij}{}^{k} \, A_k~,
\end{align}
and using this we obtain

\begin{align}
[G_{ij},G_{kl}] &= \frac{1}{2} (\delta_{il} G_{jk} + \delta_{jk} G_{il} - \delta_{ik} G_{jl} - \delta_{jl} G_{ik}) + \frac{1}{4} (\ast_{{}_7}\varphi_{ij [k}{}^{m} G_{\ell]m} - \ast_{{}_7}\varphi_{k\ell [i}{}^{m} G_{j]m})~, \notag \\
[A_i, G_{jk}] &= \frac{1}{2} (\delta_{ij} \, A_{k} - \delta_{ik} \, A_j) + \frac{1}{4} \ast_{{}_7}\varphi_{ijk}{}^{l} A_l~, \notag \\
[A_i, A_j] &= \varphi_{ij}{}^{k} \, A_k - 4 G_{ij}~.
\end{align}
Clearly, ${Spin(7)}/{G_2}$ is a homogeneous space.  As $G_2$ acts with the irreducible 7-dimensional representation on $\mathfrak{m}$,  the left-invariant metric on ${Spin(7)}/{G_2}$ is unique up to scale, therefore we may choose an ortho-normal frame $\bbe^i$ such that

\begin{align}
ds^2 = a\,  \delta_{ij} \bbl^i \bbl^j= \delta_{ij} \bbe^i \bbe^j~,
\end{align}
where $a>0$ is a constant. The left-invariant forms are

\begin{align}
\varphi = \frac{1}{3!} \, \varphi_{ijk} \, \bbe^{i}\wedge \bbe^{j}\wedge\bbe^{k}~,
\end{align}
and its dual $\ast_{{}_7}\varphi$. So the $Y$ flux can be chosen as
\begin{align} \label{Yspin7}
Y= \alpha \, \ast_{{}_7}\!\varphi~,~~ \alpha = \mathrm{const}~.
\end{align}
Using this the algebraic KSE \eqref{malgkse} can be expressed as
\begin{align}\label{Spin7G2_alg_KSE}
\left(\frac{1}{6} \alpha \left( P_1 -P_2 +P_3 - P_1 \, P_2 \, P_3 - P_2 \, P_3 + P_1 \, P_3 -  P_1 \, P_2 \right) \Gamma_z + \frac{1}{3} X \, \Gamma_x \right) \sigma_+ = \frac{1}{\ell A} \sigma_+~,
\end{align}
where $\{P_1, P_2, P_3\}= \{ \Gamma^{1245}, \Gamma^{1267}, \Gamma^{1346}\}$ are mutually commuting, hermitian Clifford algebra operators with eigenvalues $\pm 1$.
The solutions of the algebraic KSE on the eigenspaces of $\{P_1, P_2, P_3\}$ have been tabulated in table \ref{table3xxx}.
\vskip 0.5cm

\begin{table}[h]
\begin{center}
\vskip 0.3cm
 \caption{Decomposition of (\ref{Spin7G2_alg_KSE}) KSE into eigenspaces }
 \vskip 0.3cm
\begin{tabular}{|c|c|}
		\hline
		$|P_1,P_2,P_3\rangle$&  relations for the fluxes\\
		\hline
		$|+,+,+\rangle$, $|+,+,-\rangle$, $|-,+,+\rangle$, $|+,-,-\rangle$& $(-\frac{1}{6} \alpha \, \Gamma_z + \frac{1}{3} X \Gamma_{x} ) |\cdot\rangle = \frac{1}{\ell A} | \cdot \rangle$ \\
		$|-,+,-\rangle$, $|-,-,+\rangle$, $|-,-,-\rangle$&   \\
		\hline
		$|+,-,+\rangle$ & $(\frac{7}{6} \alpha \, \Gamma_z + \frac{1}{3} X \Gamma_{x} ) |\cdot\rangle = \frac{1}{\ell A} | \cdot \rangle$ \\
		\hline
		
	\end{tabular}
 \vskip 0.2cm
  \label{table3xxx}
 \end{center}
\end{table}

For backgrounds preserving    $N>16$ supersymmetries, one has  to choose the first set of solutions in  table \ref{table3xxx} and so impose the condition

\begin{align}
\frac{1}{36} \alpha^2 + \frac{1}{9} X^2 = \frac{1}{\ell^2 A^2}~.
\label{algfluxspin7}
\end{align}
However, the field equation for the warp factor $A$  \eqref{m-einst-ads} gives

\begin{align}
\frac{3}{\ell^2 A^2} = \frac{1}{3} X^2 + \frac{7}{6} \alpha^2~.
\label{feqwarpfactorspin7}
\end{align}
These two equations imply that $\alpha=0$ and so $Y=0$.

As $Y=0$, the algebraic KSE  is simplified to
\begin{align}
\Gamma_x \sigma_+ =  \frac{3}{\ell A X} \, \sigma_+~,
\label{algspin7pro}
\end{align}
and so  $\sigma_+$  lies   in one of the 8-dimensional  eigenspaces of $\Gamma_x$ provided that
 $X= \pm \frac{3}{\ell A}$.  Thus  instead of preserving 28 supersymmetries, the solution   can  be maximally supersymmetric.  Indeed this is the case as we shall now demonstrate.  The integrability
  condition of the gravitino KSE \eqref{mkse} becomes
\begin{align}
\left( R_{ijk \ell} \, \Gamma^{k \ell} -\frac{1}{18} \, X^2 \Gamma_{ij} \right) \sigma_+ = 0~.
\label{intgravspin7}
\end{align}
To investigate whether this can yield a new condition on $\sigma_+$, we find after  a direct computation using  the results of appendix B that  the Riemann tensor in the ortho-normal frame is given by
\begin{align}
R_{ijk\ell} = {9\over4} a^{-1}  ( \delta_{ik} \delta_{j\ell} - \delta_{i\ell} \delta_{jk} )~.
\end{align}
So  $S^7={Spin(7)}/{G_2}$ is equipped with the  round metric. For supersymmetric solutions, one must set $a^{-1}={1\over 81} X^2={1\over 9\ell^2 A^2}$.
In such a case, the integrability condition of the gravitino KSE  is automatically satisfied and so the solution preserves 32 supersymmetries.  This solution  is locally isometric to
the maximally supersymmetric AdS$_4\times S^7$ solution.

\subsection{\texorpdfstring{$N=24,~{SU(4)}/{SU(3)}$}{N=24, SU(4) over SU(3)}}

 As $\mathfrak{so}(6)= \mathfrak{su}(4)$, it follows from (\ref{homo167}) that the internal space  of  an AdS$_4$ solution with 24 supersymmetries is the 7-dimensional homogeneous manifold ${SU(4)}/{SU(3)}$. The geometry of this homogeneous space is  described in  Appendix~C.  The left-invariant metric can be rewritten as
\bea
ds^2= a\, \delta_{mn} \ell^m \ell^n+ b\, (\ell^7)^2= \delta_{mn}\bbe^m \bbe^n+ (\bbe^7)^2~,
\eea
where we have introduced an ortho-normal frame $\bbe^m=\sqrt{a}\, \ell^m, \bbe^7=\sqrt{b}\, \ell^7$, and  $m,n=1,\dots, 6$.
The most general left-invariant  4-form flux $Y$ can be chosen as
\begin{align}
Y =  \frac{1}{2}\, \alpha\, \omega\wedge\omega + \beta \ast_{{}_7}\!(\mathrm{Re}\,\chi) + \gamma \ast_{{}_7}\!(\mathrm{Im}\, \chi)~,
\end{align}
where  $\alpha,\beta,\gamma $ are constants and  the left-invariant 4-forms are
\bea
 \ast_{{}_7}(\mathrm{Re}\,\chi)&=& \bbe^{1367} + \bbe^{1457} + \bbe^{2357}  - \bbe^{2467}~, ~~~\omega = \bbe^{12} + \bbe^{34} + \bbe^{56}~,~~
\cr
\ast_{{}_7}(\mathrm{Im}\, \chi)&=& -\bbe^{1357} + \bbe^{1467} + \bbe^{2367} + \bbe^{2457}~,
\eea
expressed in terms of the ortho-normal frame.
Having specified the fields, it remains to solve the KSEs. For this define the mutually commuting Clifford algebra operators
\bea
&&J_1 = \cos\theta\,\Gamma^{1367} + \sin\theta\, \Gamma^{2457}~, \quad J_2 = \cos\theta\,\Gamma^{1457} + \sin\theta\,\Gamma^{2367}~, \quad
\cr
&&J_3 = \cos\theta\,\Gamma^{2357} + \sin\theta\,\Gamma^{1467}~,
\eea
with eigenvalues $\pm 1$,
where $\tan\theta=\gamma/\beta$. Then upon inserting $Y$ into the algebraic KSE \eqref{malgkse} and using the above Clifford algebra operators, we obtain
\bea
&&\Big[ -\frac{\alpha}{6} (J_1 J_2 + J_1 J_3 + J_2 J_3) ~ \Gamma_z + \frac{\sqrt{\beta^2+\gamma^2}}{6} (J_1 + J_2 + J_3 + J_1J_2J_3) ~ \Gamma_z
\cr
&&\qquad \qquad \qquad\qquad+ \frac{1}{3} X \Gamma_x\Big] \, \sigma_+ = \frac{1}{\ell A} \sigma_+~.
\label{429ccc}
\eea
The algebraic KSE \eqref{malgkse} can then be decomposed into the eigenspaces of $J_1, J_2$ and $J_3$.
 The different relations on the fluxes for all possible sets of eigenvalues of these operators are listed in   table \ref{table4xxxx}.

\begin{table}[h]
\begin{center}
\vskip 0.3cm
 \caption{Decomposition of (\ref{429ccc}) KSE into eigenspaces}
 \vskip 0.3cm

	\begin{tabular}{|c|c|}
		\hline
		$|J_1,J_2,J_3\rangle$&  relations for the fluxes\\
		\hline
		$|+,+,-\rangle$, $|+,-,+\rangle$, $|-,+,+\rangle$ & $(\frac{1}{6} \alpha \, \Gamma_z + \frac{1}{3} X \Gamma_{x} ) |\cdot\rangle = \frac{1}{\ell A} | \cdot \rangle$ \\
		$|+,-,-\rangle$, $|-,+,-\rangle$, $|-,-,+\rangle$&   \\
		\hline
		$|+,+,+\rangle$ & $[(- \frac{\alpha}{2} + \frac{2}{3} \sqrt{\beta^2+\gamma^2}) \, \Gamma_z + \frac{1}{3} X \Gamma_{x} ] |\cdot\rangle = \frac{1}{\ell A} | \cdot \rangle$ \\
		\hline
		$|-,-,-\rangle$ & $[(- \frac{\alpha}{2} - \frac{2}{3} \sqrt{\beta^2+\gamma^2}) \, \Gamma_z + \frac{1}{3} X \Gamma_{x} ] |\cdot\rangle = \frac{1}{\ell A} | \cdot \rangle$ \\
		\hline
		
	\end{tabular}
\vskip 0.2cm
  \label{table4xxxx}
 \end{center}
\end{table}

The only possibility  to obtain solutions with   $ N>16$  supersymmetries is to choose the first set of eigenspinors in table \ref{table4xxxx}. This leads to the integrability condition
\begin{align}
\frac{\alpha^2}{36} + \frac{1}{9} X^2 = \frac{1}{\ell^2 A^2}~,
\end{align}
from the remaining KSE. This together with  the warp factor field equation \eqref{m-einst-ads}
\begin{align}
\frac{1}{3} X^2 + \frac{1}{2} \alpha^2 + \frac{2}{3} (\beta^2 + \gamma^2) = \frac{3}{\ell^2 A^2}~,
\end{align}
implies
\begin{align}
\frac{5}{4} \alpha^2 + 2 (\beta^2 + \gamma^2) = 0~,
\end{align}
and so $\alpha=\beta=\gamma=0$.  Therefore   $Y=0$ and the solution is electric. As a result, the algebraic KSE \eqref{m-einst-ads} becomes
\begin{align}
\Gamma_x \sigma_+ = \frac{3}{\ell A X} \sigma_+~,
\end{align}
and so for $X=\pm 3 \ell^{-1}A^{-1}$  it admits 8 linearly independent  $\sigma_+$ solutions.  So potentially, the background is maximally supersymmetric.

It remains to investigate the gravitino KSE. First of all, we observe that for $Y=0$ the Einstein equation \eqref{m-einst-trans} along the internal space becomes
\begin{align}
R_{ij}=\frac{1}{6} X^2 \delta_{ij}~.
\end{align}
Therefore, the internal space is Einstein.  After some computation using the results in appendix C,  one finds that the homogeneous space $SU(4)/SU(3)$ is Einstein provided that    $b={9\over4}a$. In that case, the curvature of the metric in the ortho-normal frame becomes
\begin{align}
R_{ij,mn} = \frac{1}{4a} (\delta_{im} \delta_{jn} - \delta_{in} \delta_{jm})~,
\end{align}
and so the internal space is locally isometric to the round 7-sphere.  As expected from this,
the integrability condition of  the gravitino KSE \eqref{mkse}
\begin{align}
(R_{ij,mn} \Gamma^{mn} - \frac{1}{18} X^2 \Gamma_{ij}) \sigma_+ = 0~,
\label{intads4grav}
\end{align}
 has non-trivial solutions for $X^2=9 a^{-1}$, i.e. $a=\ell^2 A^2$ and $b={9\over4}\ell^2 A^2$.   With this identification of parameters,  $AdS_4\times {SU(4)}/{SU(3)}$ is locally isometric to the maximally supersymmetric $AdS_4\times S^7$ background.

To summarize there are no $AdS_4$ solutions with internal space ${SU(4)}/{SU(3)}$ which preserve  $16<N<32$ supersymmetries. However, for the choice of parameters for which
 ${SU(4)}/{SU(3)}$ is the round 7-sphere, the solution preserves 32 supersymmetries as expected.

\subsection{\texorpdfstring{$N=20$}{N=20}}
 As mentioned in the introduction, the internal space of AdS$_4$ backgrounds that preserve 20 supersymmetries admits an effective and  transitive action of a group which has Lie algebra $\mathfrak{so}(5)=\mathfrak{sp}(2)$.
The field equation for  $Y$~\eqref{Xfieldeqn} is
\begin{equation}
d\ast_{{}_7} Y = X\, Y \, .
\end{equation}
As $X$ is constant, note that for generic 4-forms $Y$ this defines a nearly-parallel $G_2$-structure on $M^7$, see e.g. \cite{Reidegeld2009} for homogeneous $G_2$ structures.  However, in what follows we shall not assume that $Y$ is generic.  In fact
in many cases, it vanishes.

Amongst the  7-dimensional compact homogeneous spaces of  (\ref{homo167}), there are three candidate internal spaces.
These are the Berger space $B^7 = Sp(2)/Sp(1)_{\mathrm{max}}$, $V_2(\bR^5) = Sp(2)/\Delta(Sp(1))$, and $J^7 = Sp(2)/Sp(1)$, corresponding to the three inequivalent embeddings of $Sp(1)$ into $Sp(2)$.
We will in the following examine each case separately, starting with the Berger space $Sp(2)/Sp(1)_{\mathrm{max}}$.

\subsubsection{\texorpdfstring{$Sp(2)/Sp(1)_{\mathrm{max}}$}{Sp(2) over Sp(1)max}}

The description of the Berger space  $B^7=Sp(2)/Sp(1)_{\mathrm{max}}$ as a   homogeneous manifold is summarized in  appendix D.  $B^7$ is diffeomeorphic to the total space of an $S^3$ bundle over $S^4$ with Euler class $\mp 10$ and first Pontryagin class $\mp16$ \cite{difftype}.  As a result   $H^4(B^7,\mathbb{Z})=\mathbb{Z}_{10}$ and $B^7$ is a rational homology 7-sphere.
 As $\mathfrak{sp}(2)= \mathfrak{so}(5)$ and $\mathfrak{sp}(1)= \mathfrak{so}(3)$, one writes $\mathfrak{so}(5)=\mathfrak{so}(3)\oplus \mathfrak{m}$ and the subalgebra $\mathfrak{so}(3)$ acts irreducibly on $\mathfrak{m}$ with the ${\bf 7}$ representation.  So  $B^7$ admits a unique
 invariant metric up to a scale and it is  Einstein. As the embedding of $\mathfrak{so}(3)$ into $\mathfrak{so}(7)$ factors through $\mathfrak{g}_2$, it also admits  an invariant 3-form $\varphi$ given in
 (\ref{g2phi}) which is unique up to a scale.  Because there is a unique invariant 3-form  $\varphi$,  $d\varphi\varpropto \ast_{{}_7}\varphi$ and $B^7$ is a nearly parallel $G_2$ manifold. Using these, we find that  the invariant fields of the theory are
\begin{equation}\label{SO5SO3_max_G_inv_metric}
ds^2= a \delta_{ij} \bbl^i \bbl^j= \delta_{ij} \bbe^i \bbe^j ~,~~~~Y={1\over 4!} \alpha  \ast_{{}_7}\varphi_{ijkm}\, \bbe^{i}\wedge \bbe^j\wedge \bbe^k\wedge \bbe^{m}~,
\end{equation}
where we have introduced the ortho-normal frame $\bbe^i=\sqrt{a}\, \bbl^i$, $\ast_{{}_7}\varphi$ is given in (\ref{g2phi}) and $a, \alpha$ are constants with $a>0$.

As the pair $(ds^2, Y)$ exhibits the same algebraic relations as that of the $Spin(7)/G_2$ case,  the algebraic KSE \eqref{Spin7G2_alg_KSE} can be solved in the same way
yielding the results of table \ref{table3xxx}. To find $N>16$ AdS$_4$ solutions, one should consider the first set of eigenspinors  of the table
which in turn imply the relation (\ref{algfluxspin7}) amongst the fluxes. This together with the field equation of the warp factor (\ref{feqwarpfactorspin7}) leads again to the
conclusion that $\alpha=0$ and so $Y=0$.

As a result of the analysis  of the algebraic KSE, so far the background can admit up to 32 supersymmetries.  It remains to investigate the solutions of the gravitino KSE.
  The curvature of $B^7$ is given by
\bea
R_{ij, km}={1\over 10 \,a} \delta_{k[i}\, \delta_{j]m}-{1\over 5 a} \ast_{{}_7}\varphi_{ijkm}+{1\over a} \delta_{\alpha\beta} k^\alpha_{ij} k^\beta_{km}~,
\eea
where $k^\alpha$ is given in appendix D.
The integrability condition of the gravitino KSE for $Y=0$ is given in (\ref{intgravspin7}).  To solve this condition, we decompose the expression into the ${\bf 7}$ and ${\bf 14}$
representations of $\mathfrak{g}_2$ using the projectors
\bea
(P^{\bf 7})^{ij}{}_{km}={1\over 3} (\delta^i_{[k} \delta^j_{m]}-{1\over2} \ast_{{}_7}\varphi^{ij}{}_{km})~,~~~(P^{\bf 14})^{ij}{}_{km}={2\over 3} (\delta^i_{[k} \delta^j_{m]}+{1\over4} \ast_{{}_7}\varphi^{ij}{}_{km})~,
\eea
and noting that $k^\alpha$ as 2-forms are in the ${\bf 14}$ representation.  The integrability condition along the ${\bf 7}$ representation gives
$X^2={81\over 5} a^{-1}$ while along the  ${\bf 14}$ representation gives that the Killing spinors must be invariant under  $\mathfrak{g}_2$.  It turns out that there are two such $\sigma_+$
spinors however taking into account the remaining projection arising from the algebraic KSE, see (\ref{algspin7pro}),  we deduce that the solution preserves 4 supersymmetries in total.  This solution   has already been
derived in \cite{Castellani:1983yg}.

\subsubsection{\texorpdfstring{${Sp(2)}/{\Delta(Sp(1))}$}{Sp(2) over Delta Sp(1)}}

The decomposition  of the Lie algebra $\mathfrak{sp}(2)= \mathfrak{so}(5)$ suitable to describe this homogeneous space can be found in appendix E.   Writing $\ell^A m_A= \ell^{ra} M_{ra}+ \ell^7 T_7$ for the left-invariant frame, $r=1,2,3$ and $a=4,5$, the most general left-invariant metric is
\bea
ds^2=\delta_{rs} g_{ab} \ell^{ra} \ell^{sb}+ a_4 (\ell^7)^2~,~~~
\eea
where $g_{ab}$ is a positive definite symmetric  $2\times 2$-matrix, $a>0$ a constant, and the left-invariant forms are generated by
\bea
 \ell^7=\ell^7~,~~~{1\over2} \delta_{rs} \epsilon_{ab} \ell^{ra}\wedge \ell^{sb}~,~~~{1\over3!} \epsilon_{rst} \ell^{ra}\wedge \ell^{sb}\wedge \ell^{tc}~.
\eea
To simplify the analysis of the geometry that follows, we note  that  without loss of generality the matrix $(g_{ab})$ can chosen to be diagonal.  To see this, perform
an orthogonal transformation $O\in SO(2)$ to bring $(g_{ab})$ into a diagonal form.  Such a transformation can be compensated with a frame rotation
\bea
\ell^{ra}\rightarrow O^a{}_b \ell^{rb}~.
\eea
Demanding that $\ell^A m_A$ is invariant implies that $M_{ra}$ has to transform as $M_{ra}\rightarrow O^b{}_a M_{rb}$.  However, it is straight forward to observe that such a
transformation is an automorphism  of $\mathfrak{so}(5)$ that preserves the decomposition (\ref{comsp2so3a}), i.e. the structure constants of the Lie algebra
remain the same. As a result, we can diagonalize the metric and at the same time use the same structure constants to calculate the geometric
quantities of the homogeneous space. Under these orthogonal transformations the first two left-invariant forms are invariant
while there is a change of basis in the space of left-invariant 3-forms.

To continue take\footnote{We have performed the analysis that follows also without taking $(g_{ab})$ to be diagonal producing the same conclusions.}  $(g_{ab})=\mathrm{diag}(a_1, a_2)$.  Then  introduce the ortho-normal frame $\bbe^7= \sqrt{a_4}\, \ell^7, \bbe^{r4}= \sqrt{a_1}\, \ell^{r4}$ and  $\bbe^{r5}= \sqrt{a_2}\, \ell^{r5}$.  In this
frame the most general left-invariant metric and $Y$ flux can be   written as
\bea
ds^2&=&\delta_{ab} \delta_{rs} \bbe^{ra} \bbe^{sb}+(\bbe^7)^2~,~~~
\cr
Y&=&\beta_1\, \bbe^7\wedge \chi_{444}+ \beta_2\, \bbe^7\wedge\chi_{445}+\beta_3\, \bbe^7\wedge\chi_{455}+ \beta_4\, \bbe^7\wedge\chi_{555}+\beta_5\, \psi~,
\eea
where $\beta_1, \beta_2, \cdots, \beta_5$ are constants,
\bea
\chi_{abc}= {1\over3!} \epsilon_{rst} \bbe^{ra}\wedge \bbe^{sb}\wedge \bbe^{tc}~,~~~\psi={1\over 2} \omega\wedge \omega~,
\eea
and
\bea
\omega={1\over2} \delta_{rs} \epsilon_{ab} \bbe^{ra}\wedge \bbe^{sb}~.
\eea
The Bianchi identity for $Y$ is automatically satisfied.  On the other hand the field equation for $Y$ in (\ref{Xfieldeqn}) yields the conditions
\bea
&&{\beta_3\over2} \sqrt{{a_2\over a_4 a_1}} - \beta_1 X=0~,~~~
-\beta_2 \sqrt{{a_2\over a_4 a_1}}+{3\beta_4 \over 2} \sqrt{{a_1\over a_4 a_2}}- \beta_2 X=0~,
\cr
&&{3\beta_1\over 2} \sqrt{{a_2\over a_4 a_1}} -\beta_3 \sqrt{{a_1\over a_2 a_4}}- \beta_3 X=0~,~~~
{\beta_2\over 2} \sqrt{{a_1\over a_4 a_2}} - \beta_4 X=0~,
\cr
&&\beta_5 \Big(X+\sqrt{{a_4\over a_1 a_2}}\Big)=0~,
\label{sp2u2beta}
\eea
where we have chosen the top form on $M^7$ as $d\mathrm{vol}= \bbe^7\wedge \chi_{444}\wedge \chi_{555}$.

Before we proceed to investigate the various cases which arise from solving the linear system (\ref{sp2u2beta}), let us consider first the case in which $F$ is electric, i.e. it is proportional to the volume form of AdS$_4$.  In such a case $\beta_1=\cdots =\beta_5=0$.  The algebraic KSE then gives
\bea
{1\over3} X \Gamma_x\sigma_+={1\over \ell A} \sigma_+~,
\eea
and the field equations  along $M^7$ imply that
\bea
R_{ij}={1\over 6} X^2 \delta_{ij}~,
\eea
and so $M^7$ is Einstein. The Einstein condition on the metric of $M^7$ requires that
\bea
a_1=a_2~,~~~a_4={3\over2} a_1~.
\eea
To investigate whether there are solutions preserving 20 supersymmetries, it remains to consider the integrability condition of the gravitino KSE (\ref{intads4grav}).
Indeed using the expressions (\ref{curvso5so3ab1}) and (\ref{curvso5so3ab2}) for the curvature of  this homogeneous space, the integrability condition along the directions $7$ and $ra$ gives $X^2=(27/8) a_1^{-1}$ while along the $ra$ and $sb$ directions requires additional
projections.  For example after taking the trace with $\delta^{ab}$ and setting $r=1$ and $s=2$,  the condition is $\Gamma^{1245}\sigma_+=\sigma_+$ which  leads to solutions
that preserve 16 supersymmetries or less, where the gamma matrices are in the ortho-normal frame and $\Gamma^{r4}=\Gamma^r, \Gamma^{r5}=\Gamma^{3+r}$.  Hence there are no $N>16$ AdS$_4$ solutions.

Next let us turn to investigate the solutions of the linear system (\ref{sp2u2beta}). The last condition implies that
\bea
\mathrm{either}\quad  \beta_5=0 \quad \mathrm{or}\quad X=-\sqrt{{a_4\over a_1 a_2}}~.
\label{sp2u2X}
\eea

To continue consider first the case that $\beta_5\not=0$.
\vskip 0.3cm
$\underline {\beta_5\not=0}$

Substituting the second equation in (\ref{sp2u2X}) into the linear system (\ref{sp2u2beta}), one finds that
\bea
\beta_3 {a_2\over 2}+ \beta_1 a_4 =0~,~~~ (a_4-a_1) \beta_3+{3\over2} a_2 \beta_1=0~,
\cr
\beta_2 {a_1\over2}+ a_4 \beta_4=0~,~~~(a_4-a_2) \beta_2+{3\over2} a_1 \beta_4=0~.
\label{loonsyst}
\eea
Now there are several cases to consider.  First suppose that the parameters of the metric $a_1,a_2,a_4$ are such that the only solutions of the
linear system above are $\beta_1=\beta_2=\beta_3=\beta_4=0$.  In such case $Y=\beta_5 \psi$ and $Y$ has the same algebraic properties as that of the $SU(4)/SU(3)$ case
with $\beta=\gamma=0$ and $\alpha=\beta_5$.  As a result, the algebraic KSE  together with the Einstein equation for the warp factor imply that $\beta_5=0$ as well and so $Y=0$.
This violates our assumption that $\beta_5\not=0$. In any case, the 4-form flux $F$ is electric which we have already investigated above and have found that such a configuration  does not admit solutions with $N>16$ supersymmetries.

Next suppose that the parameters of the metric are chosen such that
\bea
\mathrm{either} \quad \beta_1=\beta_3=0~,\quad \mathrm{or} \quad \beta_2=\beta_4=0~.
\eea
These two cases are symmetric so it suffices to consider one of the two.  Suppose that $\beta_2=\beta_4=0$ and $\beta_1, \beta_3\not=0$.  In such  a case
\bea
{3\over4} a_2^2- a_4 (a_4-a_1)=0~,
\eea
with ${3\over4} a_1^2- a_4 (a_4-a_2)\not=0$.  Setting $P_1=\Gamma^{7156}, P_2=\Gamma^{7345}$ and $P_3=\Gamma^{7264}$, the algebraic KSE can be written as
\bea
&&\Big[{1\over18} \Big(-3\beta_1 P_1P_2P_3+\beta_3 (P_1+P_2+P_3)-3\beta_5 (P_1P_2+P_1 P_3+P_2P_3) \Big) \Gamma_z
\cr
&&\qquad +{1\over 3} X \Gamma_x\Big] \sigma_+={1\over \ell A} \sigma_+~.
\label{455cccv}
\eea
As $P_1, P_2, P_3$ are commuting and have eigenvalues $\pm1$, the above algebraic equation decomposes into eigenspaces as tabulated in table \ref{table5xxxxxy}.

\begin{table}[h]
\begin{center}
\vskip 0.3cm
 \caption{Decomposition of (\ref{455cccv}) KSE into eigenspaces }
 \vskip 0.3cm
\begin{tabular}{|c|c|c|}\hline
$|P_1, P_2, P_3\rangle$ & relations for the fluxes \\ \hline
$|+,+,+\rangle$&$[ \frac{1}{6} (-\beta_1+\beta_3-3\beta_5) \Gamma_z + \frac13 X \Gamma_x ] |\cdot\rangle =  \frac{1}{\ell A} |\cdot\rangle$
\\ \hline
$|+,+,-\rangle$, $|+,-,+\rangle$, $|-,+,+\rangle$ & $[ \frac{1}{18} (3\beta_1+\beta_3+3\beta_5) \Gamma_z + \frac13 X \Gamma_x ] |\cdot\rangle =  \frac{1}{\ell A} |\cdot\rangle$ \\ \hline
$|-,-,+\rangle$, $|-,+,-\rangle$, $|+,-,-\rangle$ & $[ \frac{1}{18} (-3\beta_1-\beta_3+3\beta_5) \Gamma_z + \frac13 X \Gamma_x ] |\cdot\rangle =  \frac{1}{\ell A} |\cdot\rangle$ \\ \hline
$|-,-,-\rangle$ & $[ \frac{1}{6} (\beta_1-\beta_3-3\beta_5) \Gamma_z + \frac13 X \Gamma_x ] |\cdot\rangle =  \frac{1}{\ell A} |\cdot\rangle$ \\ \hline
	\end{tabular}
\vskip 0.2cm
  \label{table5xxxxxy}
 \end{center}
\end{table}

To find solutions with 20 supersymmetries or more, we can either choose one of the two eigenspaces with 3 linearly independent eigenspinors and both eigenspaces with a single
eigenspinor or both eigenspaces with 3 linearly independent eigenspinors. In the former case the algebraic KSE will admit 20 Killing spinors and in the latter 24 Killing spinors.

Let us first consider the case with 20 Killing spinors.  In such a case, we find that
\bea
\beta_1=\beta_3~,~~~\beta_1=3 \beta_5~,
\eea
and
\bea
{1\over36} \beta_1^2+{1\over9} X^2= {1\over\ell^2 A^2}~,
\label{algintsp2}
\eea
where we have considered the second eigenspace with 3 eigenspinors in table \ref{table5xxxxxy}. The  case where the first such eigenspace with 3 eigenspinors is chosen can be treated in a similar way.
The condition  (\ref{algintsp2}) follows as an integrability condition to the remaining algebraic KSE involving $\Gamma_z$ and $\Gamma_x$.
On the other hand, the field equation of the warp factor (\ref{m-einst-ads}) implies that
\bea
{7\over 54} \beta_1^2+{1\over 9} X^2= {1\over\ell^2 A^2}~,
\eea
which together with (\ref{algintsp2}) gives $\beta_1=0$ and so $Y=0$. The solution cannot preserve $N>16$ supersymmetries.

Next consider the case with 24 Killing spinors.  In this case, we find that
\bea
3\beta_1=-\beta_3~,
\eea
and the integrability of the remaining algebraic KSE gives
\bea
{1\over36} \beta_5^2+ {1\over9} X^2={1\over\ell^2 A^2}~.
\label{algintsp2x}
\eea
On the other hand the field equation of the warp factor (\ref{m-einst-ads}) gives
\bea
{1\over 9} X^2+{2\over 9} \beta_1^2+{1\over 6} \beta_5^2={1\over\ell^2 A^2}~.
\eea
Comparing this with (\ref{algintsp2x}), one finds that the $\beta$'s vanish and so $Y=0$. Thus there are no solutions with $N>16$ for  either $\beta_1, \beta_3$ or $\beta_2, \beta_4$  non-vanishing.

It remains to investigate the case that all $\beta_1, \dots, \beta_5\not=0$. This requires that the determinant of the coefficients of the linear system (\ref{loonsyst})
must vanish, i.e.
\bea
{3\over4} a_2^2- a_4 (a_4-a_1)=0~,~~~{3\over4} a_1^2- a_4 (a_4-a_2)=0~.
\label{2detsp2}
\eea
Taking the difference of the two equations, we find that
\bea
\mathrm{either} \quad a_1=a_2~,\quad \mathrm{or}\quad a_4={3\over4} (a_1+a_2)~.
\eea
Substituting $a_4$ above into (\ref{2detsp2}), we find that $a_1=a_2$.  So without loss of generality, we set $a_1=a_2=a$.
Then the linear system (\ref{loonsyst}) can be solved to yield
\bea
\beta_3=-3\beta_1~,~~~\beta_2=-3\beta_4~.
\eea
Setting
\bea
&&P_1=\cos\theta \Gamma^{7156}+\sin\theta \Gamma^{7234}~,~~P_2=\cos\theta \Gamma^{7345}+\sin\theta \Gamma^{7126}~,~~
\cr
&&P_3=\cos\theta \Gamma^{7264}+\sin\theta \Gamma^{7315}~,
\label{procosx}
\eea
 the algebraic KSE (\ref{malgkse}) can be rewritten as
\bea
&&\big [{1\over18}\big(\alpha P_1P_2P_3+ \alpha (P_1+P_2+P_3)-3\beta_5 (P_1P_2+P_1P_3+P_2P_3)\big) \Gamma_z
\cr
&&\qquad \qquad\qquad\qquad +{1\over3} X\Gamma_x] \sigma_+={1\over\ell A} \sigma_+~,
\label{466cccv}
\eea
where $\tan\theta=\beta_3/\beta_2$ and $\alpha=\sqrt{\beta_2^2+\beta_3^2}$.
As these Clifford algebra operations commute and have eigenvalues $\pm1$, the restrictions of this equation to the eigenspaces of $P_1, P_2$ and $P_3$ are given in  table \ref{table5xxxxxyy}.

\begin{table}[h]
\begin{center}
\vskip 0.3cm
 \caption{Decomposition of (\ref{466cccv}) KSE into eigenspaces }
 \vskip 0.3cm
\begin{tabular}{|c|c|c|}\hline
$|P_1, P_2, P_3\rangle$ & relations for the fluxes \\ \hline
$|+,+,+\rangle$&$[ \frac{1}{18} (4\alpha-9\beta_5) \Gamma_z + \frac13 X \Gamma_x ] |\cdot\rangle =  \frac{1}{\ell A} |\cdot\rangle$
\\ \hline
$|+,+,-\rangle$, $|+,-,+\rangle$, $|-,+,+\rangle$ &  \\
$|-,-,+\rangle$, $|-,+,-\rangle$, $|+,-,-\rangle$ & $[ \frac{1}{6} \beta_5 \Gamma_z + \frac13 X \Gamma_x ] |\cdot\rangle =  \frac{1}{\ell A} |\cdot\rangle$ \\ \hline
$|-,-,-\rangle$ & $[ \frac{1}{18} (-4\alpha-9\beta_5) \Gamma_z + \frac13 X \Gamma_x ] |\cdot\rangle =  \frac{1}{\ell A} |\cdot\rangle$ \\ \hline
	\end{tabular}
\vskip 0.2cm
  \label{table5xxxxxyy}
 \end{center}
\end{table}
To find solutions with 20 supersymmetries, one needs  to consider the eigenspace in table  \ref{table5xxxxxyy} with  6 eigenspinors.  In such a case the integrability
of the remaining KSE requires that
\bea
{1\over 36}\beta_5^2+{1\over9} X^2={1\over\ell^2A^2}~.
\eea
Comparing this with the field equation of the warp factor
\bea
{1\over 9}X^2 +{1\over 6} \beta^2_5+{1\over 18} (\beta_1^2+\beta_4^2)+{1\over 54} (\beta_2^2+\beta_3^2)={1\over\ell^2 A^2}~,
\eea
we find that all $\beta$'s must vanish and so $Y=0$.  Thus the flux $F$ is electric and  as we have demonstrated such background does not admit  $N>16$ AdS$_4$ supersymmetries.

\vskip 0.3cm
$\underline {\beta_5=0}$

Since the backgrounds with electric flux $F$  cannot preserve $N>16$ supersymmetries, we have to assume that at least one of the pairs $(\beta_1, \beta_3)$ and $(\beta_2, \beta_4)$
do not vanish. If either the pair $(\beta_1, \beta_3)$ or  $(\beta_2, \beta_4)$ is non-vanishing, the investigation of the algebraic KSE proceeds as in the previous case
with $\beta_5\not=0$. In particular, we find that the algebraic KSE (\ref{malgkse}) together with the field equation for the warp factor imply that all $\beta$'s vanish
and the flux $F$ is electric.  So there are no solutions preserving $N>16$ supersymmetries.

It remains to investigate the case that $\beta_1, \beta_2, \beta_3, \beta_4\not=0$.  If this is the case, the determinant of the linear system (\ref{sp2u2beta}) must vanish which
in turn implies that
\bea
-{3\over4} {a_2\over a_1a_4}+X \Big(X+\sqrt{a_1\over a_2a_4}\Big)=0~,~~~-{3\over4} {a_1\over a_2a_4}+X \Big(X+\sqrt{a_2\over a_1a_4}\Big)=0~.
\label{det2ssp22}
\eea
The solution of these equations is
\bea
\mathrm{either}\quad  a_1=a_2~,\quad \mathrm{or} \quad X=-{3\over4} {a_1+a_2\over \sqrt{a_1a_2a_4}}~.
\label{2xsp2xy}
\eea
Substituting the latter equation into (\ref{det2ssp22}), one again finds that $a_1=a_2$.  So without loss of generality we take $a_1=a_2$   in which case
\bea
\mathrm{either}\quad X={1\over 2\sqrt{a_4}}~,\quad \mathrm{or} \quad X=-{3\over 2\sqrt{a_4}}~.
\label{2xsp2xy1}
\eea
For the latter case, the linear system (\ref{sp2u2beta}) gives
\bea
\beta_3=-3\beta_1~,~~~\beta_2=-3\beta_4~.
\eea
After setting   $\beta_5=0$,  the investigation  of the algebraic KSE can be carried out as  that described in table \ref{table5xxxxxyy}. As a result after comparing with the
field equation for the warp factor, the $\beta$'s vanish and $F$ is electric.  Thus there are no solutions preserving $N>16$ supersymmetries.

\begin{table}[h]
\begin{center}
\vskip 0.3cm
 \caption{Decomposition of (\ref{474cccv}) KSE into eigenspaces }
 \vskip 0.3cm
\begin{tabular}{|c|c|c|}\hline
$|P_1, P_2, P_3\rangle$ & relations for the fluxes \\ \hline
$|+,+,+\rangle$, $|-,-,-\rangle$&$ \frac13 X \Gamma_x |\cdot\rangle =  \frac{1}{\ell A} |\cdot\rangle$
\\ \hline
$|+,+,-\rangle$, $|+,-,+\rangle$, $|-,+,+\rangle$ & $[ \frac{2}{9} \alpha \Gamma_z + \frac13 X \Gamma_x ] |\cdot\rangle =  \frac{1}{\ell A} |\cdot\rangle$ \\ \hline
$|-,-,+\rangle$, $|-,+,-\rangle$, $|+,-,-\rangle$ & $[ -\frac{2}{9} \alpha \Gamma_z + \frac13 X \Gamma_x ] |\cdot\rangle =  \frac{1}{\ell A} |\cdot\rangle$ \\ \hline
	\end{tabular}
\vskip 0.2cm
  \label{table5xxxyzz}
 \end{center}
\end{table}

It remains to investigate the case that $X=1/ (2\sqrt{a_4})$ in (\ref{2xsp2xy1}).  In this case, the linear system (\ref{sp2u2beta}) gives
\bea
\beta_1=\beta_3~,~~~\beta_2=\beta_4~.
\eea
Using the $P_1, P_2$ and $P_3$ as in (\ref{procosx}), the algebraic KSE (\ref{malgkse}) becomes
\bea
\big [{1\over18}\big(-3\alpha P_1P_2P_3+ \alpha (P_1+P_2+P_3)\big) \Gamma_z+{1\over3} X\Gamma_x] \sigma_+={1\over\ell A} \sigma_+~,
\label{474cccv}
\eea
and the solutions in the eigenspaces of $P_1,P_2$ and $P_3$ are described in table \ref{table5xxxyzz}.
To preserve $N>16$ supersymmetries, one has to consider either one of the eigenspaces with 3 eigenspinors and the eigenspace with 2 eigenspinors or both of the eigenspaces with 3
eigenspinors.  In either case, one finds that all $\beta$'s vanish and so $Y=0$.  Then $F$ is electric  and such solutions do not preserve $N>16$ supersymmetries.
Therefore we conclude that the homogenous space ${Sp(2)}/{\Delta(Sp(1))}$ does not give rise to AdS$_4$ backgrounds with $N>16$.

\subsubsection{\texorpdfstring{${Sp(2)}/{Sp(1)}$}{Sp(2) over Sp(1)}}

The geometry of this homogeneous space is  described in appendix E where the definition of the generators of the algebra and expressions
for the curvature and invariant forms can be found.  A left-invariant frame is  $\ell^A m_A= \ell^a W_a+ \ell^r T^{(+)}_r$, where $a=1,\dots, 4$ and $r=1,2,3$.  Then the most general left-invariant metric is
\bea
ds^2= a \delta_{ab} \bbl^a \bbl^b+ g_{rs} \bbl^{r} \bbl^s~,
\label{mettrr}
\eea
where  $a>0$ is a constant and  $(g_{rs})$ is any constant $3\times 3$  positive definite symmetric matrix.

To simplify the computations that follow, it is convenient to use the covariant properties of the decomposition of $\mathfrak{sp}(2) = \mathfrak{so}(5)$ as described in (\ref{so5so3aa})  to restrict the number of parameter in the metric. In particular, observe that the decomposition  (\ref{so5so3aa}) remains invariant
under the transformation of the generators
\bea
T_r^{(+)}\rightarrow O_r{}^s T_s^{(+)}~,~~~W_a\rightarrow U_a{}^b W_b~,~~~T_r^{(-)}\rightarrow  T^{(-)}_r~,
\eea
where $O\in SO(3)$ and $U\in Spin(3)\subset SO(4)$  defined as
\bea
O_r{}^s I^{(+)}_s=  U I^{(+)}_r U^{-1}~,
\eea
as $I^{(+)}_r$ are  the gamma matrices of the Majorana spinor
representation of $\mathfrak{so}(3)$ on $\bR^4= \bC^2\oplus \bar\bC^2$. Furthermore notice that
$U I^{(-)}_r U^{-1}=I^{(-)}_r$ as $U$ is generated by $I^{(+)}_r$ which commute with all $I^{(-)}_s$.
The orthogonal rotations $O$ act on the matrix $(g_{rs})$ as $g\rightarrow O g O^{-1}$.  As $(O, U)$
is an automorphism of $\mathfrak{so}(5)$ which leaves the decomposition (\ref{so5so3aa})
invariant,  we can use $O$ to put the matrix $(g_{rs})$ into diagonal form.  So from now on without loss
of generality, we set $(g_{rs})=\mathrm{diag}(b_1, b_2, b_3)$ with $b_1, b_2, b_3>0$, see also \cite{ziller}.

The left-invariant 4-forms are generated by
\bea
\psi={1\over 4!}\epsilon_{abcd} \ell^a\wedge \ell^b\wedge \ell^c\wedge \ell^d~,~~~\rho_{rs}={1\over2} \epsilon_{rpq} \ell^p\wedge \ell^q \wedge I^{(+)}_s~,
\eea
where
\bea
I^{(+)}_s={1\over2} (I^{(+)}_s)_{ab}\, \ell^a\wedge \ell^b~.
\eea
Therefore the 4-form flux $Y$ can  be chosen as
\bea
Y= \alpha \psi + \beta^{rs} \rho_{rs}~,
\eea
where $\alpha$ and $\beta^{rs}$ are constants.
Then it is straightforward to find  that the Bianchi identity $dY=0$ implies that
\bea
\beta^{rs}=\beta^{sr}~.
\eea
Furthermore define $\sigma={1\over 3!}  \epsilon_{rst} \ell^r\wedge \ell^s\wedge \ell^t$ and choose as top form $d\mathrm{vol}=a^2 \sqrt{b_1 b_2b_3}\, \sigma\wedge \psi$. Then the field equation for $Y$, $d\ast_{{}_7}Y=XY$, gives the linear system
\bea
\sum_{r=1}^3 b_r \beta^{rr}= \sqrt{b_1b_2b_3}\, X \alpha~,~~~{\alpha\over2} {\sqrt{b_1b_2b_3}\over a^2}-{1\over3} {\sum_{r=1}^3 b_r \beta^{rr}\over \sqrt{b_1b_2b_3}}={X\over 3} \beta
\cr
\big( b_r \beta^{rs}+\beta^{rs} b_s-{2\over 3} \delta^{rs} \sum_{t=1}^3 b_t \beta^{tt}\big)=\sqrt{b_1b_2b_3} X (\beta^{rs}-{1\over3} \delta^{rs} \beta)~,
\label{lastls}
\eea
where there is no summation over the indices $r$ and $s$ on the left-hand side of the last equation and $\beta=\delta_{rs} \beta^{rs}$.

Before we proceed to investigate the solutions of the linear system, notice that if $\beta^{rs}=0$,
then $\alpha=0$ and so $F$ is electric.  The supersymmetry preserved by these
solutions will be investigated later.  As we shall demonstrate  such solutions cannot preserve more than 16
supersymmetries.

Furthermore writing $Y=\alpha \psi+ Y_\beta$, where $Y_\beta=\beta^{rs} \rho_{rs}$, the field equation
of the warp factor in (\ref{m-einst-ads}) can be written as
\bea
{1\over 9} X^2+{1\over18} \alpha^2 a^{-4} +{1\over 432} (Y_\beta)^2={1\over \ell^2 A^2}~.
\label{k1ller}
\eea
As we shall demonstrate, the compatibility of this field equation with the algebraic
KSE rules out the existence of $N>16$ backgrounds.

Returning to the solutions of (\ref{lastls}), let us focus on $\beta^{rs}$ with $r\not=s$. There are several cases
to consider.

\vskip0.3cm
\underline{Either $\beta^{rs}\not=0$ for all $r\not=s$ or $\beta^{rs}=0$ for all $r\not=s$ }

If $\beta^{rs}$,  $r\not=s$, are all non-vanishing, the last equation in  (\ref{lastls}) implies that

\bea
b_1=b_2=b_3~,~~~ X=2{b_1\over \sqrt{b_1b_2b_3}}~.
\eea
  As a result, the metric is invariant under  $SO(3)$ and this can be used  to
bring $\beta^{rs}$ into diagonal form.  Of course $(\beta^{rs})$ is also diagonal if
$\beta^{rs}=0$ for all $r\not=s$.

So without loss of generality, we can assume   that  $(\beta^{rs})$
is diagonal. Setting
\bea
J_1= \Gamma^{6714}~,~~~J_2=\Gamma^{6723}~,~~~J_3=\Gamma^{7524}~,
\eea
where all gamma matrices are in the ortho-normal basis and $\{\Gamma^i\}=\{\Gamma^a, \Gamma^{4+r}\}$,
the algebraic KSE can be written as
\bea
&&\Big({1\over6} \Big[-\alpha a^{-2} J_1 J_2+ {a^{-1} \over \sqrt{b_1 b_2b_3}}  \big(\sqrt{b_1}\beta^{11} (J_1+J_2)
+\sqrt{b_2} \beta^{22} J_3 (1+ J_1J_2)
\cr &&
+ \sqrt{b_3} \beta^{33}
J_3 (J_1+J_2)\big)\Big] \Gamma_z
+{1\over3} X \Gamma_x\Big) \sigma_+={1\over \ell A} \sigma_+~.
\label{493cccv}
\eea
The decomposition of the algebraic KSE into the eigenspaces of the commuting Clifford algebra
operators $J_1, J_2, J_3$ is  illustrated in table \ref{table5omega}.
\begin{table}[ht]
\begin{center}
\vskip 0.3cm
 \caption{Decomposition of (\ref{493cccv}) KSE into eigenspaces }
 \vskip 0.3cm
\begin{tabular}{|c|c|c|}\hline
$|J_1, J_2, J_3\rangle$ & relations for the fluxes \\ \hline
$|+,+,\pm\rangle$ & $\begin{array}{r@{}l@{}} \big( \frac{1}{6} [-\alpha a^{-2} & + 2 {a^{-1} \over \sqrt{b_1 b_2b_3}} (\sqrt{b_1}
 \beta^{11}\pm\sqrt{b_2} \beta^{22} \\ & \hspace*{1cm} \pm\sqrt{b_3} \beta^{33})] \Gamma_z + \frac13 X \Gamma_x \big) |\cdot\rangle =  \frac{1}{\ell A} |\cdot\rangle \end{array}$ \\ \hline
 $|+,-,\pm\rangle$ $|-,+,\pm\rangle$& $\left( \frac{1}{6} \alpha a^{-2} \Gamma_z + \frac13 X \Gamma_x \right) |\cdot\rangle =  \frac{1}{\ell A} |\cdot\rangle$
 \\
 \hline
$|-,-,\pm\rangle$ & $\begin{array}{r@{}l@{}} \big( \frac{1}{6} [-\alpha a^{-2} & + 2 {a^{-1} \over \sqrt{b_1 b_2b_3}} (-\sqrt{b_1}
 \beta^{11}\pm \sqrt{b_2} \beta^{22} \\ & \hspace*{1cm} \mp\sqrt{b_3} \beta^{33})] \Gamma_z + \frac13 X \Gamma_x \big) |\cdot\rangle =  \frac{1}{\ell A} |\cdot\rangle \end{array}$
 \\
 \hline
	\end{tabular}
\vskip 0.2cm
  \label{table5omega}
 \end{center}
\end{table}

To construct $N>16$ solutions, we have to include the eigenspace
with  four eigenspinors.  The integrability condition of the remaining KSE described in table
\ref{table5omega} gives
\bea
{1\over36} \alpha^2 a^{-4} +{1\over 9} X^2={1\over \ell^2 A^2}~.
\label{intconomega}
\eea
Comparing (\ref{intconomega}) with the field equation for the warp factor (\ref{k1ller}), we
find that $\alpha=\beta^{rs}=0$.  Therefore $Y=0$ and so $F$ is electric.

\vskip 0.3cm

{\underline{ $\beta^{12}, \beta^{13}\not=0$ and  $\beta^{23}=0$}}

As the other two cases for which either $\beta^{13}=0$ or $\beta^{12}=0$ with  the rest of the components non-vanishing can be treated in a similar way,  we take without loss of generality that $\beta^{23}=0$ and $\beta^{12}, \beta^{13}\not=0$.  In such a case, the last condition in (\ref{lastls}) gives
\bea
X={b_1+b_2\over \sqrt{b_1 b_2b_3}}~,~~~b_2=b_3~.
\eea
The metric is invariant under an $SO(2)\subset SO(3)$ symmetry which acts with the vector
representation on the vector  $(\beta^{12}, \beta^{13})$ and leaves the form
of $(\beta^{rs})$ invariant.  As a result up to an $SO(2)$ transformation, we can set $\beta^{13}=0$ as well.
Furthermore, if $b_1\not= b_2$, the diagonal terms in the last condition in (\ref{lastls})  give
\bea
\beta^{11}=-\beta^{22}=-\beta^{33}~.
\eea
On the other hand if $b_1=b_2$ the analysis reduces to that of the previous case.
Therefore for $b_1\not= b_2$, $Y$ can be written as
\bea
Y=\alpha \psi+\beta^{11} (\rho_{11}-\rho_{22}-\rho_{33})+ \beta^{12} (\rho_{12}+\rho_{21})~.
\eea
Introducing  the  Clifford algebra operators
\bea
J_1=\cos\theta \Gamma^{6714}+\sin\theta \Gamma^{6724}~,~~~J_2=\cos\theta \Gamma^{5724}-\sin\theta \Gamma^{5714}~,~~~J_3=\Gamma^{1234}~,
\label{SO5SO3Aprojectors1}
\eea
where $\tan\theta=\beta^{12}/\beta^{11}$,
the algebraic KSE can be written as
\begin{multline}
\Big({1\over6} \Big[\alpha a^{-2} J_3+\frac{a^{-1}}{\sqrt{b_1} b_2} (\sqrt{b_2}\beta^{11} J_1  J_2 (1 - J_3) \\
 + \sqrt{(\beta^{11})^2+(\beta^{12})^2} (\sqrt{b_1} J_1 + \sqrt{b_2}J_2)(1 - J_3) )\Big] \Gamma_z
+{1\over3} X \Gamma_x\Big) \sigma_+={1\over \ell A} \sigma_+ \; .
\label{493cccvi}
\end{multline}
The decomposition of the algebraic KSE into the eigenspaces of the commuting Clifford algebra
operators $J_1, J_2, J_3$ is  illustrated in table \ref{table5omega1}.

\begin{table}[ht]
\begin{center}
\vskip 0.3cm
 \caption{Decomposition of (\ref{493cccvi}) KSE into eigenspaces }
 \vskip 0.3cm
\begin{tabular}{|c|c|c|}\hline
$|J_1, J_2, J_3\rangle$ & relations for the fluxes \\ \hline
$|\pm,+,-\rangle$ & $\begin{array}{r@{}l@{}} \Big( \frac{1}{6} [-\alpha a^{-2} & +\frac{2 a^{-1}}{\sqrt{b_1} b_2} (\pm \sqrt{b_2} \beta^{11}  \\ & \hspace*{-1cm} +\sqrt{(\beta^{11})^2+(\beta^{12})^2} (\pm\sqrt{b_1}+\sqrt{b_2}))] \Gamma_z + \frac13 X \Gamma_x \Big) |\cdot\rangle =  \frac{1}{\ell A} |\cdot\rangle \end{array}$ \\ \hline
 $|+,\pm,+\rangle$ $|-,\pm,+\rangle$& $\left( \frac{1}{6} \alpha a^{-2} \Gamma_z + \frac13 X \Gamma_x \right) |\cdot\rangle =  \frac{1}{\ell A} |\cdot\rangle$
 \\
 \hline
$|\pm,-,-\rangle$ & $\begin{array}{r@{}l@{}} \Big( \frac{1}{6} [-\alpha a^{-2} & +\frac{2 a^{-1}}{\sqrt{b_1} b_2} (\mp \sqrt{b_2} \beta^{11} \\ & \hspace*{-1cm} +\sqrt{(\beta^{11})^2+(\beta^{12})^2} (\pm \sqrt{b_1}-\sqrt{b_2}))] \Gamma_z + \frac13 X \Gamma_x \Big) |\cdot\rangle =  \frac{1}{\ell A} |\cdot\rangle \end{array}$
 \\
 \hline
	\end{tabular}
\vskip 0.2cm
  \label{table5omega1}
 \end{center}
\end{table}

To construct solutions preserving more than 16 supersymmetries, we have to include the eigenspace
with  four eigenspinors leading again to the integrability condition (\ref{intconomega}).
Comparing again with the field equations of the warp factor (\ref{k1ller}), we deduce
that $F$ is electric.

\vskip 0.3cm
$\underline {\beta^{13}=\beta^{23}=0 \;\; \text{but $\beta^{12} \not=0$}}$

All three cases for which only  one of the three off-diagonal components of $(\beta^{rs})$ is non-zero  can be treated symmetrically. So without loss of generality,  one can  take  $\beta^{13}=\beta^{23}=0$ but $\beta^{12} \not=0$.
In this case, the last equation in (\ref{lastls}) has four branches of solutions depending on the choice of the  $b_1$, $b_2$ and $b_3$ components of the metric.
\begin{enumerate}
\item $b_1=b_2=b_3=b$\newline The last equation in (\ref{lastls}) then implies $X=2/\sqrt{b}$ and the aforementioned residual $SO(3)$ symmetry can be used to put $\beta^{rs}$ to be diagonal.
\item $b_1=b_2$, $b_2 \not= b_3$\newline The last equation in (\ref{lastls}) then implies $X=2/\sqrt{b_3}$ and $\beta^{33}=0$. The aforementioned residual $SO(2)$ symmetry can be used to put $\beta^{rs}$ to be diagonal.
\item $b_2 \not= b_3$, $b_1 + b_2 = 2 b_3$\newline The last equation in (\ref{lastls}) then implies  $X=(b_1+b_2)/\sqrt{b_1 b_2 b_3}$ and $\beta^{11}=\beta^{22}=0$.
In such a case, $Y$ reads
\bea
Y=\alpha \psi + \beta^{33} \rho_{33} + \beta^{12} (\rho_{12}+\rho_{21}) \; .
\eea
Choosing
\bea\label{SO5SO3Acase43projectors}
J_1=\Gamma^{1457}~,~~~J_2=\Gamma^{2467}~,~~~J_3=\Gamma^{1234} \; ,
\eea
the algebraic KSE can be written as
\begin{multline}
\Big({1\over6} \Big[\alpha a^{-2} J_3+\frac{a^{-1}}{\sqrt{b_1 b_2 b_3}} \left(\sqrt{b_3} \beta^{33} J_1 J_2 (1-J_3) \right. \\ \left. - \beta^{12} (\sqrt{b_2}J_1-\sqrt{b_1}J_2) (1-J_3)\right)\Big] \Gamma_z
+{1\over3} X \Gamma_x\Big) \sigma_+={1\over \ell A} \sigma_+ \; .
\label{SO5SO3AKSEcase43}
\end{multline}
The decomposition of the algebraic KSE into the eigenspaces of $J_1, J_2, J_3$ is  illustrated in table \ref{SO5SO3AKSEcase43KSEtable}.
\begin{table}[ht]
\begin{center}
\vskip 0.3cm
 \caption{Decomposition of (\ref{SO5SO3AKSEcase43}) KSE into eigenspaces }
 \vskip 0.3cm
\begin{tabular}{|c|c|c|}\hline
$|J_1, J_2, J_3\rangle$ & relations for the fluxes \\ \hline
$|\pm,+,-\rangle$ & $\begin{array}{r@{}l@{}} \Big( \frac{1}{6} [-\alpha a^{-2} & +\frac{2 a^{-1}}{\sqrt{b_1 b_2 b_3}} (\pm \sqrt{b_3} \beta^{33}  \\ & -\beta^{12} (\pm\sqrt{b_2}-\sqrt{b_1}))] \Gamma_z + \frac13 X \Gamma_x \Big) |\cdot\rangle =  \frac{1}{\ell A} |\cdot\rangle \end{array}$ \\ \hline
 $|+,\pm,+\rangle$ $|-,\pm,+\rangle$& $\left( \frac{1}{6} \alpha a^{-2} \Gamma_z + \frac13 X \Gamma_x \right) |\cdot\rangle =  \frac{1}{\ell A} |\cdot\rangle$
 \\
 \hline
$|\pm,-,-\rangle$ & $\begin{array}{r@{}l@{}} \Big( \frac{1}{6} [-\alpha a^{-2} & +\frac{2 a^{-1}}{\sqrt{b_1 b_2 b_3}} (\mp \sqrt{b_3} \beta^{33}  \\ & -\beta^{12} (\pm\sqrt{b_2}+\sqrt{b_1}))] \Gamma_z + \frac13 X \Gamma_x \Big) |\cdot\rangle =  \frac{1}{\ell A} |\cdot\rangle \end{array}$
 \\
 \hline
	\end{tabular}
\vskip 0.2cm
  \label{SO5SO3AKSEcase43KSEtable}
 \end{center}
\end{table}
Again the eigenspace with four eigenspinors has to be included in the construction of $N>16$ backgrounds.  As a result, this leads to the integrability condition (\ref{intconomega}) which together with the warp factor field equation (\ref{k1ller}) imply that $F$ is electric.

\item $b_1 \not= b_2$, $b_1 + b_2 \not= 2 b_3$\newline The last equation in (\ref{lastls}) then implies
    \bea
    X=(b_1+b_2)/\sqrt{b_1 b_2 b_3}~,~~~ \beta^{11}=-\beta^{22}=\beta^{33} (2 b_3 - b_1 - b_2)/(b_1 - b_2)~.
    \eea
     In such a case, $Y$ reads
\bea
Y=\alpha \psi + \beta^{11} (\rho_{11}-\rho_{22}+\frac{b_1 - b_2}{2 b_3 - b_1 - b_2} \rho_{33})+ \beta^{12} (\rho_{12}+\rho_{21}) \; .
\eea
With the choice of commuting Clifford algebra operators as in~\eqref{SO5SO3Aprojectors1}, the algebraic KSE can be written as
\begin{multline}
\Big({1\over6} \Big[\alpha a^{-2} J_3+\frac{a^{-1}}{\sqrt{b_1 b_2 b_3}} \big(\frac{(b_1-b_2)\sqrt{b_3}}{b_1+b_2-2 b_3} \beta^{11} J_1 J_2 (1-J_3) \\ +\sqrt{(\beta^{11})^2+(\beta^{12})^2} (\sqrt{b_1} J_1 + \sqrt{b_2}J_2) (1-J_3) \big)\Big] \Gamma_z
+{1\over3} X \Gamma_x\Big) \sigma_+={1\over \ell A} \sigma_+ \; .
\label{SO5SO3AKSEcase44}
\end{multline}
The decomposition of the algebraic KSE into the eigenspaces of $J_1, J_2, J_3$ is  illustrated in table \ref{SO5SO3AKSEcase44KSEtable}.

To construct $N>16$ solutions, we again have to include the eigenspace
with  four eigenspinors which leads to the integrability condition (\ref{intconomega}).
Comparing with the warp factor field equation (\ref{k1ller}), we again deduce
that $F$ is electric.

\end{enumerate}

\begin{table}[ht]
\begin{center}
\vskip 0.3cm
 \caption{Decomposition of (\ref{SO5SO3AKSEcase44}) KSE into eigenspaces }
 \vskip 0.3cm
\begin{tabular}{|c|c|c|}\hline
$|J_1, J_2, J_3\rangle$ & relations for the fluxes \\ \hline
$|\pm,+,-\rangle$ & $\begin{array}{r@{}l@{}} \Big( \frac{1}{6} [-\alpha a^{-2} & +\frac{2 a^{-1}}{\sqrt{b_1 b_2 b_3}} (\pm \frac{(b_1-b_2)\sqrt{b_3}}{b_1+b_2-2 b_3} \beta^{11}  \\ & \hspace*{-1cm} +\sqrt{(\beta^{11})^2+(\beta^{12})^2} (\pm\sqrt{b_1}+\sqrt{b_2}))] \Gamma_z + \frac13 X \Gamma_x \Big) |\cdot\rangle =  \frac{1}{\ell A} |\cdot\rangle \end{array}$ \\ \hline
 $|+,\pm,+\rangle$ $|-,\pm,+\rangle$& $\left( \frac{1}{6} \alpha a^{-2} \Gamma_z + \frac13 X \Gamma_x \right) |\cdot\rangle =  \frac{1}{\ell A} |\cdot\rangle$
 \\
 \hline
$|\pm,-,-\rangle$ & $\begin{array}{r@{}l@{}} \Big( \frac{1}{6} [-\alpha a^{-2} & +\frac{2 a^{-1}}{\sqrt{b_1} b_2} (\mp \frac{(b_1-b_2)\sqrt{b_3}}{b_1+b_2-2 b_3} \beta^{11} \\ & \hspace*{-1cm} +\sqrt{(\beta^{11})^2+(\beta^{12})^2} (\pm \sqrt{b_1}-\sqrt{b_2}))] \Gamma_z + \frac13 X \Gamma_x \Big) |\cdot\rangle =  \frac{1}{\ell A} |\cdot\rangle \end{array}$
 \\
 \hline
	\end{tabular}
\vskip 0.2cm
  \label{SO5SO3AKSEcase44KSEtable}
 \end{center}
\end{table}

It remains to investigate the number of supersymmetries preserved by the solutions for which $F$ is electric.
For this, one has to investigate the integrability condition of the gravitino KSE (\ref{intads4grav}).
Using the expression for the curvature of metric in (\ref{curvsp2sp1aa})-(\ref{curvsp2sp1ad}) and requiring that the solution preserves $N>16$, we find that
\bea
\delta^{ca} \delta^{db} (I^{(-)}_r)_{ab}(R_{cd,mn} \Gamma^{mn} - \frac{1}{18} X^2 \Gamma_{cd}) \sigma_+ = 0~,
\eea
implies that
\bea
a-{1\over 8} \delta^{rs} g_{rs}-{1\over18} a^2\, X^2=0~.
\label{ina}
\eea
Next requiring again that $N>16$, one finds that the condition
\bea
\delta^{ca} \delta^{db} (I^{(+)}_r)_{ab}(R_{cd,mn} \Gamma^{mn} - \frac{1}{18} X^2 \Gamma_{cd}) \sigma_+ = 0~.
\eea
gives that
\bea
&&\delta^{pq} g_{pq} \epsilon_{rst}-{1\over2} a^{-1} \epsilon_t{}^{pq} g_{pr} g_{qs}- 2 g_{tp} \epsilon^p{}_{rs}=0~,
\cr
&&-{3\over4} g_{rs}+{1\over8} \delta^{pq} g_{pq} \delta_{rs}+ a\delta_{rs}- {1\over18} a^2\, X^2 \delta_{rs}=0~.
\label{inb}
\eea
Substituting (\ref{ina}) into the second equation in (\ref{inb}), one finds after a bit of analysis that
\bea
b_1=b_2=b_3~.
\eea
Setting  $b=b_1=b_2=b_3$ and substituting this back into (\ref{ina}) and (\ref{inb}), one deduces that
\bea
2a=b~,~~~X^2=9 b^{-1}~.
\eea
As $X^2=9 \ell^{-2} A^{-2}$, we have $b=\ell^2 A^2$ and $a=(1/2)\ell^2 A^2$.
The rest of the integrability condition is satisfied without further conditions. So every solutions that preserves $N>16$ supersymmetries is maximally supersymmetric
and so locally isometric to $AdS_4\times S^7$.

One can confirm this result  by investigating the Einstein equation (\ref{m-einst-trans}). As  all  solutions  with electric $F$  are Einstein $R^{(7)}_{ij}=(1/6) X^2 \delta_{ij}$, it suffices to identify
the left-invariant metrics on ${Sp(2)}/{Sp(1)}$ that are Einstein.  There are two Einstein metrics \cite{ziller, gibbons} on ${Sp(2)}/{Sp(1)}$ given by
\bea
X^2=9 b^{-1}~,~~~2a=b~,~~~b_1=b_2=b_3=b~,
\eea
and
\bea
X^2={81\over 25} b^{-1}~,~~~2a=5b~,~~~b_1=b_2=b_3=b~,
\eea
where the first one is the round metric on $S^7$, see also \cite{jensen}.  The second one does not give $N>16$ supersymmetric solutions.

\newsection{Conclusions}

We have classified up to local isometries all warped AdS$_4$ backgrounds with the most general allowed fluxes in 10- and 11-dimensional supergravities that preserve $N>16$ supersymmetries.
We have demonstrated that up to an overall scale, the only solutions that arise are  the maximally supersymmetric solution $AdS_4\times S^7$ of 11-dimensional
supergravity \cite{fr, duffpopemax}  and the $N=24$ solution $AdS_4\times\mathbb{CP}^3$  of IIA supergravity \cite{nillsonpope}.  These two solutions are related via dimensional reduction  along
the  fibre of the Hopf fibration $S^1\rightarrow S^7\rightarrow \mathbb{CP}^3$.

The assumption we have made  to prove these results is that either the solutions are smooth and the internal space is compact without boundary or that the even part $\mathfrak{g}_0$ of
the Killing superalgebra of the backgrounds decomposes as $\mathfrak{g}_0=\mathfrak{so}(3,2)\oplus \mathfrak{t}_0$. In fact these two assumptions are equivalent for $N>16$  AdS$_4$
backgrounds.  It may be possible to weaken these assumptions but they cannot be removed altogether.  This is because in such a case additional
solutions will exist.  For example the maximally supersymmetric $AdS_7\times S^4$ solution of 11-dimensional supergravity \cite{townsend} can be re-interpreted as a maximally supersymmetric  warped
AdS$_4$ solution. However in such case the ``internal'' 7-dimensional manifold $M^7$ is not compact and the even subalgebra of the Killing superalgebra $\mathfrak{g}_0$ does not decompose as $\mathfrak{so}(3,2)\oplus \mathfrak{t}_0$.

We have identified all AdS$_4$ backgrounds up to a local isometry. Therefore, we have specified all the local geometries of the internal spaces $G/H$  of these solutions.  However the  possibility remains that there are more solutions which arise via  additional discrete identifications   $Z\backslash G/H$, where $Z$ is a discrete subgroup of $Z\subset G$.  The  $AdS_4\times Z\backslash G/H$ solutions will preserve at most as many supersymmetries as the  $AdS_4\times G/H$ solutions.  As in IIB and massive IIA supergravities there are no $N>16$  $AdS_4\times G/H$ solutions, there are no $N>16$  $AdS_4\times Z\backslash G/H$ solutions either.   In  IIA theory, the possibility remains that there can be $AdS_4\times Z \backslash \mathbb{CP}^3$ solutions with 24 and 20 supersymmetries. In $D=11$ supergravity as $AdS_4\times S^7$ preserves 32 supersymmetries, there may be $AdS_4\times Z\backslash S^7$ solutions preserving 28, 24 and 20 supersymmetries.  Such solutions have been used in the context of AdS/CFT in \cite{abjm}. A systematic investigation of all possible $N>16$  $AdS_4\times Z\backslash G/H$ backgrounds
will involve the identification of all discrete subgroups of $G$. The relevant groups here are $SU(4)$
and $\mathrm{Spin}(8)$, see e.g. \cite{hanany} for an exposition of discrete subgroups of $SU(4)$ and references therein.

 It is clear from our results on AdS$_4$ backgrounds that supersymmetric AdS solutions which preserve $N>16$ supersymmetries in 10- and 11-dimensions are severely restricted.
Consequently there are few gravitational duals for superconformal theories with a large number of supersymmetries which have distinct local geometries.  For example, the
  superconformal theories of \cite{ferrara, aharony, garcia} have  gravitational duals which are locally isometric to  the $AdS_5\times S^5$ maximally supersymmetric background as there are no distinct local AdS$_5$ geometries that preserve strictly 24 supersymmetries \cite{ads5clas}. In general our
results also suggest that there may not be a large number of backgrounds that preserve $N>16$ supersymmetries in 10- and 11-dimensional supergravities.
So it is likely that all these solutions can be found in the future.

\section*{Acknowledgments}

The authors would like to express a special thanks to the Mainz Institute for Theoretical Physics (MITP) for its hospitality and support during the MITP Topical Workshop ``Geometry, Gravity and Supersymmetry'' (GGSUSY2017).
The work of ASH is supported by the German Science Foundation (DFG) under the Collaborative Research Center (SFB) 676 ``Particles, Strings and the Early Universe''. GP is partially supported from the  STFC rolling grant ST/J002798/1.

\setcounter{section}{0}\setcounter{equation}{0}

\appendix{Notation and conventions}

Our  conventions for forms are as follows. Let $\omega$ be a k-form, then
\bea
\omega=\frac{1}{k!} \omega_{i_1\dots i_k} dx^{i_1}\wedge\dots \wedge dx^{i_k}~,~~~\omega^2_{ij}= \omega_{i\ell_1\dots \ell_{k-1}} \omega_{j}{}^{\ell_1\dots \ell_{k-1}}~,~~~
\omega^2= \omega_{i_1\dots i_k} \omega^{i_1\dots i_k}~.
\eea
We also define
\bea
{\slashed\omega}=\omega_{i_1\dots i_k} \Gamma^{i_1\dots i_k}~, ~~{\slashed\omega}_{i_1}= \omega_{i_1 i_2 \dots i_k} \Gamma^{i_2\dots i_k}~,~~~\slashed{\gom}_{i_1}= \Gamma_{i_1}{}^{i_2\dots i_{k+1}} \omega_{i_2\dots i_{k+1}}~,
\eea
where the $\Gamma_i$ are the Dirac gamma matrices.

The inner product $\langle\cdot, \cdot\rangle$ we use on the space of spinors is that for which space-like gamma matrices are Hermitian while time-like gamma
matrices are anti-hermitian, i.e. the Dirac spin-invariant inner product is $\langle\Gamma_0\cdot, \cdot\rangle$. The norm $\parallel\cdot\parallel=\sqrt {\langle\cdot, \cdot\rangle}$ is taken with respect to $\langle\cdot, \cdot\rangle$, which is positive definite. For more details on our conventions
see \cite{mads, iiaads, iibads}.

\appendix{Homogeneous and symmetric spaces}

In the following section we shall collect some useful properties of homogeneous spaces which have facilitated our analysis of AdS$_4$ backgrounds. A more detailed review can  be found in e.g. \cite{kobayashi, muellerhossein}.

Consider the left coset space $M=G/H$, where $G$ is a compact connected semisimple Lie group $G$ which acts  effectively from the left on $M=G/H$ and $H$ is a closed Lie subgroup of $G$. Let us denote the Lie algebras of $G$ and $H$ with  $\mathfrak{g}$ and $\mathfrak{h}$, respectively. As there is always an invariant inner product on  $\mathfrak{g}$, it can be used to take the orthogonal complement of  $\mathfrak{h}$ in  $\mathfrak{g}$ and so
\begin{align}
\mathfrak{g}=\mathfrak{h} \oplus \mathfrak{m}~.
\end{align}
Denote   the generators of  $\mathfrak{h}$ with $h_\alpha$, $\alpha=1,2,..., \dim{\mathfrak{h}}$  and a basis in $\mathfrak{m}$ as  $m_A$, $A=1,..., \dim{\mathfrak{g}}-\dim{\mathfrak{h}}$. In this basis,  the brackets of the Lie algebra $\mathfrak{g}$ take the following form

\bea\label{commutation}
&&[h_\alpha, h_\beta] = f_{\alpha\beta}{}^\gamma \, h_\gamma~,~~~
[h_\alpha, m_A] = f_{\alpha A}{}^B \, m_B~,
\cr
&&[m_A,m_B] = f_{AB}{}^C \, m_C + f_{AB}{}^\alpha \, h_\alpha~.
\eea
If $f_{AB}{}^C=0$, that is $[\mathfrak{m},\mathfrak{m}] \subset \mathfrak{h}$, the space is symmetric.

Let $g: U\subset G/H\rightarrow G$  be a local section of the coset. The decomposition of the Maurer-Cartan form in components along  $\mathfrak{h}$ and  $\mathfrak{m}$ is
\begin{align}
g^{-1} dg = \bbl^A \, m_A + \Omega^\alpha \, h_\alpha~,
\label{MC}
\end{align}
which  defines a local left-invariant frame $\bbl^A$  and a canonical left-invariant connection  $\Omega^\alpha$ on  $ G/H$. The curvature and torsion of the canonical connection are
\bea\label{dei}
&&R^\alpha \equiv d\Omega^\alpha+\frac12 f_{\beta\gamma}{}^\alpha \Omega^\beta\wedge \Omega^\gamma=-\frac12 f_{BC}{}^\alpha \bbl^B\wedge \bbl^C~,
\cr
&&T^A\equiv d\bbl^A+f_{\beta C}{}^A \Omega^\beta\wedge \bbl^C=-\frac12 f_{BC}{}^A \bbl^B\wedge \bbl^C~,
\eea
respectively, where the equalities follow after  taking the exterior derivative of (\ref{MC}) and  using (\ref{commutation}).  If $G/H$ is symmetric, then
the torsion vanishes.

A left-invariant   p-form $\omega$ on $ G/H$ can be written as
\begin{align}
\omega = \frac{1}{p!} \, \omega_{A_1 ... A_p} \, \bbl^{A_1} \wedge ... \wedge \bbl^{A_p}~,
\end{align}
where the components   $\omega_{A_1...A_p}$ are constant and satisfy
\begin{align}\label{hinvariance}
f_{\alpha[A_1}{}^B \, \omega_{A_2...A_p]B} =0~.
\end{align}
The latter condition is required for invariance under the right action of $H$ on $G$. All left-invariant forms are  parallel with respect to the canonical connection.

It remains to describe the metrics of $G/H$ which are  left-invariant.  These are written as
\bea
ds^2=g_{AB}\, \bbl^A \bbl^B~,
\eea
where the components $g_{AB}$ are constant and satisfy
\begin{align}\label{ginvar}
f_{\alpha A}{}^C \, g_{BC} + f_{\alpha B}{}^C \, g_{AC} = 0~.
\end{align}
For symmetric spaces, the canonical connection coincides with the Levi-Civita connection of invariant metrics. So all non-vanishing  left-invariant forms are harmonic and
represent non-trivial elements in the de Rham cohomology of  $G/H$.
However if $G/H$ is strictly homogeneous this is not the case since the canonical connection has non-vanishing torsion.

Suppose $G/H$ is homogeneous and equipped with an invariant metric $g$. To describe the results of the paper, it is required to find the Levi-Civita connection of $g$ and  its curvature.
Let $\Phi$ be the Levi-Civita connection in the left-invariant frame. As the difference of two connections is a tensor, we set
\bea
\Phi{}^A{}_B= \Omega^\alpha f_{\alpha B}{}^A+\bbl^C Q_{C,}{}^A{}_B~.
\eea
As $\Phi$ is metric and torsion free, we have

\begin{align}\label{levi-civ}
\Phi_{AB} + \Phi_{BA} &= 0~,\notag \\
d\bbl^A + \Phi^A{}_B \, \wedge \, \bbl^B &=0~.
\end{align}
These equations can be solved for $Q$ to find that
\begin{align}
\Phi^A{}_B =  \Omega^\alpha\, f_{\alpha B}{}^A + \frac{1}{2} \left( g^{AD} \, f_{DB}{}^E \, g_{CE} +  g^{AD} \, f_{DC}{}^E \, g_{BE} + f_{CB}{}^A\right) \, \bbl^C~.
\end{align}
In turn the Riemann  curvature 2-form $R^A{}_B$ is
\begin{align}
R^A{}_B = \frac{1}{2} \left( Q_{C,}{}^A{}_{E} \, Q_{D,}{}^E{}_{B} - Q_{D,}{}^A{}_{E} \, Q_{C,}{}^E{}_{B} - Q_{E,}{}^A{}_{B} \, f_{CD}{}^E - f_{CD}{}^\alpha \, f_{\alpha B}{}^A \right) \, \bbl^C \wedge \bbl^D~.
\end{align}
This  is required for the investigation of the gravitino KSE.  Note that the expression for $\Phi^A{}_B$ is considerably  simplified whenever the coset space is naturally reductive because the structure constants $f_{ABC}=f_{AB}{}^E \, g_{CE}$ are then skew symmetric.

\appendix{\texorpdfstring{$\mathfrak{su}(k)$}{su(k)}}

Here we shall collect some  formulae that are useful in understanding the homogeneous spaces that admit a transitive action of a group with Lie algebra $\mathfrak{su}(k)$.
A basis over the reals of anti-hermitian $k\times k$ traceless complex matrices is
\bea
(M_{ab})^c{}_d=\frac12 (\delta_a{}^c \delta_{bd}-\delta_b{}^c \delta_{ad})~,~~~(N_{ab})^c{}_d=\frac{\nu(ab)}{2} i (\delta_a{}^c \delta_{bd}+\delta_b{}^c \delta_{ad}-\frac{2}{k} \delta_{ab} \delta_c{}^d)~,
\eea
where $\nu(ab)$ is a normalization factor and $a,b, c, d=1,\dots, k$. The trace of these matrices yields an invariant  inner product on $\mathfrak{su}(k)$. In particular
the non-vanishing traces are
\bea
&&\mathrm{tr}( M_{ab} M_{a'b'})=-\frac12 (\delta_{aa'} \delta_{bb'}- \delta_{ab'} \delta_{ba'})~,~~~
\cr
&&\mathrm{tr}( N_{ab} N_{a'b'})=-\frac{\nu(ab) \nu(a'b')}{2}(\delta_{aa'} \delta_{bb'}+ \delta_{ab'} \delta_{ba'}-\frac{2}{k} \delta_{ab} \delta_{a'b'})~.
\eea
It is customary to choose the normalization factors $\nu$  such that all generators have the same length. In such a case, they will depend on $k$.  However in what follows, it is  more convenient to choose $\nu=1$.
The Lie brackets of $\mathfrak{su}(k)$ are
\bea
[M_{ab}, M_{a'b'}]&=&\frac12 (\delta_{ba'} M_{ab'}+ \delta_{ab'} M_{ba'}- \delta_{aa'} M_{bb'}-\delta_{bb'} M_{aa'})~,
\cr
[M_{ab},  N_{a'b'}]&=&\frac12  (\delta_{ba'} N_{ab'}- \delta_{ab'} N_{ba'}- \delta_{aa'} N_{bb'}+\delta_{bb'} N_{aa'})~,
\cr
[N_{ab},  N_{a'b'}]&=&-\frac12 (\delta_{ba'} M_{ab'}+ \delta_{ab'} M_{ba'}+ \delta_{aa'} M_{bb'}+\delta_{bb'} M_{aa'})~.
\label{sukbra}
\eea
We shall proceed to describe the homogeneous spaces in (\ref{homo166}) and (\ref{homo167}) that admit a transitive $SU(k)$ action.

\subsection{$M^k=\mathbb{CP}^{k-1}=SU(k)/S(U(k)\times U(1))$}

To describe the  $\mathbb{CP}^{k-1}$ homogeneous space, we set
\bea
\mathfrak{h}=\mathfrak{s}(\mathfrak{u}(k-1)\oplus \mathfrak{u}(1))=\bR\langle M_{rs}, N_{rs}, N_{kk})\rangle ~,~~~\mathfrak{m}=\bR\langle M_{rk}, N_{sk}\rangle~,
\eea
where $r, s=1,\dots, k-1$.
  The brackets of the Lie subalgebra  $\mathfrak{s}(\mathfrak{u}(k-1)\oplus \mathfrak{u}(1))$ can be read off from those   in (\ref{sukbra}) while those involving
  elements of $\mathfrak{m}$ are
\bea
[M_{rk}, M_{sk}]=- \frac12 M_{rs}~,~~~[M_{rk},  N_{sk}]=\frac12   N_{rs}-\frac12 \delta_{rs} N_{kk}~,~~~
[N_{rk},  N_{sk}]=-\frac12  M_{rs}~,
\label{sukbra2}
\eea
and
\bea
[M_{rs}, M_{tk}]&=&\frac12 (\delta_{ts} M_{rk}- \delta_{tr} M_{sk})~,~~~
[M_{rs},  N_{tk}]=\frac12  (\delta_{ts} N_{rk}- \delta_{tr} N_{sk})~,
\cr
[ N_{rs}, M_{tk}]&=&\frac12  (\delta_{ts} N_{rk}+ \delta_{tr} N_{sk})~,~~~ [N_{rs},  N_{tk}]=-\frac12 (\delta_{ts} M_{rk}+ \delta_{tr} M_{sk})~,
\cr
[N_{kk},  M_{sk}]&=& - N_{rk} ~,~~~    [N_{kk},  N_{rk}]= M_{rk}~.
\label{sukbra3}
\eea
The left-invariant frame is $\bbl^A m_A= \bbl^r M_{rk}+ \bbl^{\tilde r} N_{rk}$. The most general left-invariant metric can be expressed as
\bea
ds^2=a\, (\delta_{rs}  \bbl^r \bbl^s+ \delta_{\tilde r\tilde s}  \bbl^{\tilde r} \bbl^{\tilde s})~,~~~
\eea
where $a>0$ is a constant.  The left-invariant forms of $\mathbb{CP}^{k-1}$ are generated by the (K\"ahler) 2-form
\bea
\omega= a\, \delta_{r\tilde s} \ell^r\wedge \ell^{\tilde s}~.
\eea
 The non-vanishing components of the curvature of the metric in the ortho-normal frame are
\bea
R_{rs, pq}&=&-\frac{1}{4a} (\delta_{rq} \delta_{sp}- {1\over a}\delta_{rp} \delta_{sq}) ~,~~~R_{rs, \tilde p\tilde q}=-\frac{1}{4a} (\delta_{r\tilde q} \delta_{s\tilde p}- {1\over a}\delta_{r\tilde p} \delta_{s\tilde q})~,~~~
\cr
R_{r\tilde s, p\tilde q}&=& \frac{1}{4a} (\delta_{r\tilde q} \delta_{\tilde sp}+ \delta_{rp} \delta_{\tilde s\tilde q})+\frac{1}{2a} \delta_{r\tilde s} \delta_{p\tilde q}~,~~~R_{\tilde r\tilde s, \tilde p\tilde q}=-\frac{1}{4 a} (\delta_{\tilde r\tilde q} \delta_{\tilde s\tilde p}- \delta_{\tilde r\tilde p} \delta_{\tilde s\tilde q})~.
\eea
This expression of the curvature matches that in (\ref{cp3curv}) for $\mathbb{CP}^3$ up to an overall scale.

\subsection{$M^k=SU(k)/SU(k-1)$}

Next let us turn to the $SU(k)/SU(k-1)$ homogeneous space.  The embedding of $\mathfrak{su}(k-1)=\bR\langle M_{rs}^{(k-1)}, N_{rs}^{(k-1)}\rangle $, where $r,s=1,\dots, k-1$,  into $\mathfrak{su}(k)=\bR\langle M_{ab}^{(k)}, N_{ab}^{(k)}\rangle$ is given by
\bea
M^{(k-1)}_{rs}=M^{(k)}_{rs}~,~~~N^{(k-1)}_{rs}=N^{(k)}_{rs}+{1\over k-1} \delta_{rs} N^{(k)}_{kk}~.
\eea
As $\mathfrak{m}=\bR\langle  M_{rk}^{(k)}, N_{sk}^{(k)},  N^{(k)}_{kk}\rangle$, the  (non-vanishing) commutators involving elements of $\mathfrak{m}$ are
\bea
&&[M^{(k)}_{rk}, M^{(k-1)}_{sk}]=- \frac12 M^{(k)}_{rs}~,~~~[M^{(k)}_{rk},  N^{(k)}_{sk}]=\frac12   N^{(k-1)}_{rs}-\frac{k}{2(k-1)} \delta_{rs} N^{(k)}_{kk}~,~~~
\cr
&&[N^{(k)}_{rk},  N^{(k)}_{sk}]=-\frac12  M^{(k-1)}_{rs}~,
\label{sukbra2x}
\eea
and
\bea
[M^{(k-1)}_{rs}, M^{(k)}_{tk}]&=&\frac12 (\delta_{ts} M^{(k)}_{rk}- \delta_{tr} M^{(k)}_{sk})~,~~~
[M^{(k-1)}_{rs},  N^{(k)}_{tk}]=\frac12  (\delta_{ts} N^{(k)}_{rk}- \delta_{tr} N^{(k)}_{sk})~,
\cr
[ N^{(k-1)}_{rs}, M^{(k)}_{tk}]&=&-{1\over k-1} \delta_{rs} N_{tk}^{(k)}+\frac12  (\delta_{ts} N^{(k)}_{rk}+ \delta_{tr} N^{(k)}_{sk})~,~~~
 \cr
 [N^{(k-1)}_{rs},  N^{(k)}_{tk}]&=&{1\over k-1} \delta_{rs} M^{(k)}_{tk}-\frac12 (\delta_{ts} M^{(k)}_{rk}+ \delta_{tr} M^{(k)}_{sk})~,
\cr
[N^{(k)}_{kk},  M^{(k)}_{rk}]&=& - N^{(k)}_{rk} ~,~~~    [N^{(k)}_{kk},  N^{(k)}_{rk}]= M^{(k)}_{rk}~.
\label{sukbra3x}
\eea
Setting $\bbl^A m_A= \hat\bbl^r M^{(k)}_{rk}+ \hat\bbl^{\tilde r} N^{(k)}_{rk}+\hat\bbl^0 N^{(k)}_{kk}$ for the left-invariant frame, a direct computation reveals that the most general invariant metric is
\bea
&&ds^2=a\, (\delta_{rs}  \hat\bbl^r \hat\bbl^s+ \delta_{\tilde r\tilde s}  \hat\bbl^{\tilde r} \hat\bbl^{\tilde s})+ b (\hat\bbl^0)^2~,~~~
\eea
where $a,b>0$ are constants. Moreover the left-invariant 2- and 3-forms  for $k=4$ are generated by
\bea
\hat\omega=\delta_{r\tilde s} \hat\bbl^r\wedge \hat\bbl^{\tilde s}~,~~~\hat\bbl^0\wedge \hat\omega~,~~~\mathrm{Re}\, \hat\chi~,~~~\mathrm{Im}\, \hat\chi~,
\eea
and their duals, where
\bea
 \hat\chi=\frac{1}{3!}\,\epsilon_{rst} (\hat\bbl^r+ i\hat\bbl^{\tilde r}) \wedge  (\hat\bbl^s+ i\hat\bbl^{\tilde s}) \wedge  (\hat\bbl^t+ i\hat\bbl^{\tilde t})~,
\eea
is the holomorphic (3,0)-form.

 However for convenience, we re-label the indices of the left-invariant frame as
$ \ell^{2r-1}= \hat\ell^r, \ell^{2r}=\hat\ell^{\tilde r}, \ell^7=\hat\ell^0$, $r=1,2,3$  in which case the left-invariant metric can be rewritten as
\bea
ds^2= a\, \delta_{mn} \ell^m \ell^n+ b\, (\ell^7)^2= \delta_{mn}\bbe^m \bbe^n+ (\bbe^7)^2~,
\eea
where we have introduced an ortho-normal frame $\bbe^m=\sqrt{a}\, \ell^m, \bbe^7=\sqrt{b}\, \ell^7$, and  $m,n=1,\dots, 6$.
Note also that up to an overall scale, the left-invariant 2- and 3-forms can be re-written in terms of the ortho-normal frame.  In particular, we have
\begin{align}
\omega &= \bbe^{12} + \bbe^{34} + \bbe^{56}~,~~ \bbe^7\wedge \omega~,~~~\mathrm{Re}\, \chi~,~~~\mathrm{Im}\, \chi~,
\end{align}
where
\bea
 \chi= (\bbe^1+ i\bbe^{ 2}) \wedge  (\bbe^3+ i\bbe^4) \wedge  (\bbe^5+ i \bbe^6)~.
\eea
We shall use this ortho-normal basis to solve the KSEs for this internal space.

\appendix{The Berger space \texorpdfstring{$B^7={Sp(2)}/{Sp(1)_{\text{max}}}$}{Sp(2) over Sp(1)max}}

To describe the geometry of the Berger space $B^7$, one  identifies   the vector representation ${\bf 5}$ of  $\mathfrak{so}(5)=\mathfrak{sp}(2)$ with the symmetric trace-less
representation of $\mathfrak{so}(3)=\mathfrak{sp}(1)$ and then decomposes the adjoint representation of $\mathfrak{so}(5)$ in $\mathfrak{so}(3)$ representations as
${\bf 10}={\bf 3}\oplus {\bf 7}$, where ${\bf 7}$ is the symmetric traceless representation of $\mathfrak{so}(3)$ constructed with three copies of the vector
representation. As a result $\mathfrak{so}(5)=\mathfrak{so}(3)\oplus \mathfrak{m}$, where $\mathfrak{so}(3)$ and $\mathfrak{m}$ are identified with the 3-dimensional
and 7-dimensional representations, respectively.

This decomposition can be implemented as follows.  Consider the basis $W_{ab}$, $a,b,c,d=1,\dots,5$,
\bea
(W_{ab}){}^c{}_d=\delta_a^c \delta_{bd}- \delta_b^c \delta_{ad}~,
\eea
in $\mathfrak{so}(5)$ leading to the commutators
\bea
[W_{ab}, W_{a'b'}]= (\delta_{ba'} W_{ab'}+\delta_{ab'} W_{ba'}- \delta_{aa'} W_{bb'}-  \delta_{bb' } W_{aa'})~.
\eea
Then re-write each basis element using the ${\bf 5}$ representation $\mathfrak{so}(3)$  as
$W_{rs, tu}$, where $r, s,t,u=1,2,3$.  Decomposing  this into $\mathfrak{so}(3)$ representations, one  finds that
\bea
W_{rs,tu}&=&O_{ru} \delta_{st}+O_{su} \delta_{rt}+O_{rt} \delta_{su}+O_{st} \delta_{ru}
\cr~~~&&+ \epsilon^p{}_{st} S_{pru}+\epsilon^p{}_{rt} S_{psu}+\epsilon^p{}_{su} S_{prt}+\epsilon^p{}_{ru} S_{pst}~,
\eea
where  $O\in \mathfrak{so}(3)$ and $S\in \mathfrak{m}$.  Using this one can proceed to describe the homogeneous space $B^7$.  However, this decomposition
does not automatically reveal the $G_2$ structure which  is necessary in the analysis of the supersymmetric solutions.  Instead, we shall follow an adaptation \cite{difftype} of the
description in \cite{Castellani:1983yg} and  \cite[Appendix A.1]{Haupt:2015wdq}.  For this use the inner product
\bea
\langle W_{ab}, W_{a'b'}\rangle=-{1\over2} \mathrm{tr} ( W_{ab} W_{a'b'} )~,
\eea
which is $\mathfrak{so}(5)$ invariant and the basis $W_{ab}$, $a<b$, is ortho-normal.  In this basis, the structure constants of $\mathfrak{so}(5)$ are skew-symmetric.
Then identify the $\mathfrak{so}(3)$ subalgebra of $\mathfrak{so}(5)$ with the span of the ortho-normal vectors
\bea
h_1&=& {1\over\sqrt5} (-W_{12}-W_{34}+\sqrt3 W_{35})~,~~~h_2={1\over\sqrt5} (-W_{13}+W_{24}+\sqrt{3} W_{25})~,~~~
\cr
h_3&=&{1\over\sqrt5} (-2W_{14}+W_{23})~.
\eea
We choose the subspace $\mathfrak{m}$ to be orthogonal to $\mathfrak{so}(3)$ and an ortho-normal basis in $\mathfrak{m}$  introduced as
\bea
&&m_1={1\over2\sqrt5} (4W_{12}- W_{34}+\sqrt3 W_{35})~,~~~m_2={1\over2\sqrt5} ( 4W_{13}+W_{24}+\sqrt3 W_{25})~,~~
\cr &&
m_3={1\over \sqrt5} (-W_{14}-2 W_{23})  ~,~~~m_4={1\over2} (\sqrt 3W_{34}+W_{35})~,~~~m_5={1\over2} (\sqrt 3 W_{24}-W_{25})~,~~~
\cr
&&
m_6=W_{15}~,~~~m_7=W_{45}~.
\eea
Then it is straightforward to  show that
\bea
[h_\alpha, h_\beta]={1\over\sqrt 5} \epsilon_{\alpha\beta}{}^\gamma h_\gamma~,~~~[h_\alpha, m_i]= k_{\alpha i}{}^j m_j~,~~~[m_i, m_j]={1\over \sqrt 5} \varphi_{ij}{}^k m_k+ k_{ij}{}^\alpha h_\alpha~,~~~
\eea
where $\varphi$ is given in (\ref{g2phi}), the indices are raised and lowered with the flat metric and
\bea
k^1&=&-{3\over 2\sqrt 5} m_2\wedge m_3-{\sqrt 3\over 2} m_2\wedge m_6-{\sqrt 3\over 2} m_3\wedge m_5+{2\over  \sqrt 5} m_4\wedge m_7+{1\over 2 \sqrt 5} m_5\wedge m_6~,
\cr
k^2&=& {3\over 2\sqrt 5} m_1\wedge m_3-{\sqrt 3\over 2}m_1\wedge m_6-{\sqrt 3\over 2} m_3\wedge m_4-{1\over 2 \sqrt 5} m_4\wedge m_6+{2\over  \sqrt 5} m_5\wedge m_7~,
\cr
k^3&=&-{3\over 2\sqrt 5} m_1\wedge m_2-{\sqrt 3\over 2} m_1\wedge m_5-{\sqrt 3\over 2} m_2\wedge m_4+{1\over 2 \sqrt 5} m_4\wedge m_5+{2\over  \sqrt 5} m_6\wedge m_7~.
\nonumber
\eea
So $f_{ij}{}^k= {1\over \sqrt 5} \varphi_{ij}{}^k$ and the Jacobi identities imply that $\varphi$ is invariant under the representation of $\mathfrak{so}(3)$ on $\mathfrak{m}$.  Therefore the  embedding
of $\mathfrak{so}(3)$ in $\mathfrak{so}(7)$ defined by  $(k^1, k^2, k^3)$  factors through $\mathfrak{g}_2$.  This is useful in the analysis of the gravitino KSE.

\setcounter{subsection}{0}\setcounter{equation}{0}

\appendix{ $\mathfrak{so}(5)=\mathfrak{sp}(2)$ } \label{so5append3x}

To describe the various homogeneous spaces that we are using which admit a transitive action of a group with Lie algebra $\mathfrak{so}(5)=\mathfrak{sp}(2)$,
 choose a basis  in $\mathfrak{so}(5)$ as
\begin{align}\label{gen-sona}
(M_{\tilde a\tilde b})_{\tilde c\tilde d} = \frac12 ( \delta_{\tilde a\tilde c} \, \delta_{\tilde b\tilde d} - \delta_{\tilde a\tilde d}\, \delta_{\tilde b\tilde c}),
\end{align}
in  $\mathfrak{sp}(2) = \mathfrak{so}(5)$, where $M_{\tilde a\tilde b},{} \tilde a,\tilde b=1,..,5$.  The commutators are
\begin{align}\label{son-commuta}
[M_{\tilde a\tilde b}, M_{\tilde a'\tilde b'}] = \frac{1}{2} (\delta_{\tilde a\tilde b'} M_{\tilde b\tilde a'} + \delta_{\tilde b\tilde a'} M_{\tilde a\tilde b'} - \delta_{\tilde b\tilde b'} M_{\tilde a\tilde a'} - \delta_{\tilde a\tilde a'} M_{\tilde b\tilde b'})~.
\end{align}
In what follows, we shall describe various decompositions $\mathfrak{so}(5)=\mathfrak{h}\oplus \mathfrak{m}$ for different  choices of a subalgebra $\mathfrak{h}$ and
summarize some of their algebraic and geometric properties that we are using in this work.

\subsection{$M^6=Sp(2)/U(2)$}

The subalgebra $\mathfrak{h}$ and $\mathfrak{m}$ are spanned as
 \bea
 \mathfrak{u}(2) = \mathfrak{u}(2) \equiv \bR\, \langle T_r,T_{7}\rangle =\bR\, \langle \frac12 \epsilon_r{}^{st} M_{st},  M_{45}\rangle~,
 \label{decsp2u2aaa}
 \eea
 and
\begin{align}
\mathfrak{m}=\bR\,\langle M_{ra} \rangle = \bR\,\langle M_{r4},M_{r5},\rangle~,
\end{align}
respectively,  where $r,s,t=1,2,3$ and  $a, b, c, \ldots = 4, 5$.
 In this basis the non-vanishing commutators are
\bea
&&[T_r, T_s]= -\frac12 \epsilon_{rs}{}^t T_t~,~~[T_r, M_{sa}]=-\frac12 \epsilon_{rs}{}^t M_{ta}~,~~~[T_7, M_{ra}]=-\frac12\epsilon_{ab} M_{rb}~,~~~
\cr
&&[M_{ra}, M_{sb}]=-\frac12\delta_{ab} \epsilon_{rs}{}^t T_t-\frac12\delta_{rs} \epsilon_{ab} T_7~.
\label{comsp2so3a}
\eea
Clearly this is a symmetric coset space admitting an invariant metric
\bea
ds^2=a\,\delta_{rs} \delta_{ab} \bbl^{ra} \bbl^{sb}=\delta_{rs} \delta_{ab} {\bf e}^{ra} {\bf e}^{sb}~,
\label{sp2u2metra}
\eea
where $a>0$ is a constant, and $\bbl^{ra}$ and ${\bf e}^{ra}=\sqrt{a}\,\bbl^{ra} $ are the left-invariant and ortho-normal frames, respectively.
 The curvature of the symmetric space in the ortho-normal frame is
 \bea
 R_{ra\,sb, tc\, ud}=\frac{1}{4a} (\delta_{rt} \delta_{su} -\delta_{ru} \delta_{st}) \delta_{ab} \delta_{cd}+ \frac{1}{4 a}\delta_{rs} \delta_{tu} \epsilon_{ab} \epsilon_{cd}~,
 \label{curvsp2u2}
 \eea
which is instrumental in the investigation of the gravitino KSE in section 3.5.1.

\subsection{$M^6={Sp(2)}/({Sp(1)\times U(1)})$}

Viewing the elements of $Sp(2)$  as quaternionic $2\times2$ matrices,   ${Sp(1)\times U(1)}\subset Sp(1)\times Sp(1)$ is embedded in $Sp(2)$ along the diagonal. To describe this embedding choose
a basis in $\mathfrak{sp}(2)=\mathfrak{so}(5)$ as in (\ref{gen-sona}) and set
\bea
T^{(\pm)}_r=\frac{1}{2} \varepsilon^{rst} M^{st} \pm M^{r4}~,~~~, W_a=\sqrt{2} M_{a5}~,
\eea
where  $r=1,2,3$ and now $a=1,\dots 4$.
In terms of this basis,  the non-vanishing commutators of $\mathfrak{sp}(2)$ are
\bea
[T^{(\pm)}_r, T^{(\pm)}_s]&=&-\epsilon_{rs}{}^ t T^{(\pm)}_t~,~~~[T^{(\pm)}_r, W_a]=\frac12 (I^{(\pm)}_r)^b{}_a W_b~,~~
\cr
[W_a, W_b]&=&-\frac12\big( (I^{(+)}_r)_{ab} T^{(+)}_r+(I^{(-)}_r)_{ab} T^{(-)}_r\big)~,
\label{so5so3aa}
\eea
where
\bea
(I^{(\pm)}_r)^4{}_s=\mp \delta_{rs}~,~~~(I^{(\pm)}_r)^s{}_4=\pm \delta^s{}_r~,~~(I^{(\pm)}_r)^s{}_t=\epsilon_{rst}~.
\label{Imatrpm}
\eea
Observe that $(I^{(\pm)}_r)$ are bases in the spaces of (anti-)self-dual forms in $\bR^4$ and that
\bea
I^{(\pm)}_r I^{(\pm)}_s=-\delta_{rs} {\bf 1}-\epsilon_{rst} I^{(\pm)}_t~.
\eea
The subalgebra $\mathfrak{h}$ and $\mathfrak{m}$ are spanned as
\bea
\mathfrak{h}=\mathfrak{sp}(1)\oplus \mathfrak{u}(1)=\bR\langle T^{(-)}_r, T^{(+)}_3\rangle~,
\eea
 and
 \bea
 \mathfrak{m}=\bR\langle W_a, T^{(+)}_1, T^{(+)}_2\rangle~,
 \eea
 respectively.
Introducing the left-invariant frame,  $\bbl^A m_A= \bbl^a W_a+ \bbl^{\underline r} T^{(+)}_{\underline r}$, where ${\underline r}=1,2$, the left-invariant metric can be written
as
\bea
ds^2= a\, \delta_{ab} \bbl^a \bbl^b+ b \,\delta_{{\underline r}{\underline s}} \bbl^{\underline r}\bbl^{\underline s} = \delta_{ab} \bbe^a \bbe^{b} + \delta_{{\underline r}{\underline s}} \bbe^{{\underline r}} \bbe^{{\underline s}}~,
\eea
where $a,b>0$ and we have introduced the ortho-normal frame $\bbe^a=\sqrt{a}\, \bbl^a $, $\bbe^{{\underline r}}=\sqrt{b}\, \bbl^{\underline r}$.

The curvature of this metric in the ortho-normal frame is
\begin{align}
R_{ab,cd} &= \left( \frac{1}{2a} - \frac{3b}{16a^2} \right) ( \delta_{ac} \, \delta_{bd} - \delta_{ad} \, \delta_{bc}) + \frac{3b}{16a^2} \left( (I^{(+)}_3)_{ab}  (I^{(+)}_3)_{cd} - (I^{(+)}_3)_{a[b} (I^{(+)}_3)_{cd]} \right)~, \notag\\
R_{a\underline{r}, b \underline{s}} &= \frac{b}{16a^2} \delta_{ab}\delta_{{\underline r}{\underline s}} + \left( \frac{1}{4a} - \frac{b}{16a^2} \right) \epsilon_{{\underline r}{\underline s}} (I^{(+)}_3)_{ab}~, \notag\\
R_{ab,\underline{r}\underline{s}} &= \left( \frac{1}{2a} - \frac{b}{8a^2} \right)\epsilon_{{\underline r}{\underline s}} (I^{(+)}_3)_{ab}, \quad R_{\underline{rs},\underline{tu}} = \frac{1}{b} \epsilon_{{\underline r}{\underline s}} \epsilon_{{\underline t}{\underline u}}~.
\label{curvsp2sp1u1}
\end{align}
We shall use these expressions in the investigation of the gravitino KSE in section 3.5.2.

\subsection{$M^7={Sp(2)}/{\Delta(Sp(1))}$}

The decomposition  of the Lie algebra $\mathfrak{sp}(2) = \mathfrak{so}(5)$ suitable for the  description of  this homogeneous space is as in (\ref{decsp2u2aaa}) but now $\mathfrak{h}$ and $\mathfrak{m}$ are spanned as
\bea
\mathfrak{h}= \bR\langle T_r\rangle~,~~~\mathfrak{m}=\bR\langle M_{ra}, T_7\rangle~,
\eea
respectively, where $r=1,2,3$ and $a=4,5$.
Introducing the left-invariant frame as $\ell^A t_A= \ell^{ra} M_{ra}+ \ell^7 T_7$, the left-invariant metric is
\bea
ds^2=\delta_{rs} g_{ab} \ell^{ra} \ell^{sb}+ a_4 (\ell^7)^2~,~~~
\eea
where $(g_{ab})$ is a symmetric constant positive definite $2\times 2$ matrix and $a_4>0$ is a constant.  The curvature of this metric in the left-invariant frame  is
\bea
R_{pc\, qd,}{}^{ra}{}_{sb}&=&-{1\over16} a_4^{-1} \delta_p^r \delta_{qs} g^{ae} ((\Delta g)_{ec}- a_4 \epsilon_{ec}) ((\Delta g)_{db}+ a_4 \epsilon_{db})
\cr
&&+{1\over16} a_4^{-1} \delta_q^r \delta_{ps} g^{ae} ((\Delta g)_{ed}- a_4 \epsilon_{ed}) ((\Delta g)_{cb}+ a_4 \epsilon_{cb})
\cr
&&+{1\over8} \epsilon_{cd} \delta_{pq} \delta^r_s g^{ae} \epsilon_{eb} (\delta^{t_1t_2} g_{t_1t_2}-a_4)-{1\over4} \delta_{cd} \delta^a_b ( \delta_{ps} \delta^r_q- \delta_{qs} \delta_p^r)~,
\label{curvso5so3ab1}
\eea
and
\bea
R_{7\, ar,}{}^7{}_{bs}&=&{1\over16} a_4^{-1} ((\Delta g)_{ad}+ a_4 \epsilon_{ad}) g^{de} \epsilon_{eb} (\delta^{t_1t_2} g_{t_1t_2}-a_4) \delta_{rs}
\cr
&&-{1\over8} a_4^{-1} \epsilon_a{}^d ((\Delta g)_{db}+a_4 \epsilon_{db}) \delta_{rs}~,
\label{curvso5so3ab2}
\eea
where
\bea
(\Delta g)_{ab}= \epsilon_a{}^d g_{db}+\epsilon_b{}^d g_{da}~,
\eea
 $(g^{ab})$ is the inverse matrix of $(g_{ab})$ and the indices of $\epsilon$ are raised and lowered with $\delta_{ab}$.
The Ricci tensor again in the left-invariant frame is
\bea
R_{ra\, sb}&=&\big[{a_4^{-1} \over 16} g^{dc} (\Delta g)_{da} (\Delta g)_{cb}-{1\over 16} g^{dc}  (\Delta g)_{cb} \epsilon_{da}
\cr
&&
+{1\over 16} g^{cd} \epsilon_{ca} \epsilon_{db}
(\delta^{t_1t_2} g_{t_1t_2}-2a_4)+{a_4^{-1} \over 16} (\Delta g)_{ad} g^{dc} \epsilon_{cb}\, \delta^{t_1t_2} g_{t_1t_2}
\cr
&&- {a_4^{-1}\over 8} \epsilon_a{}^d (\Delta g)_{db}+{5\over8} \delta_{ab}\big]\delta_{rs}~,
\eea
and
\bea
R_{77}=-{3\over8} {a_4\over \det g} (\delta^{t_1t_2} g_{t_1t_2}-a_4)+{3\over8} a_4 \delta_{ab} g^{ab}-{3\over8} \epsilon_a{}^d (\Delta g)_{db} g^{ab}~.
\eea
%
It is straightforward to compute the Ricci tensor for $(g_{ab})$ diagonal. This concludes the summary of the geometry for this homogeneous space.

\subsection{$M^7=Sp(2)/Sp(1)$}

The decomposition  of the Lie algebra $\mathfrak{sp}(2) = \mathfrak{so}(5)$ suitable for the  description of this homogeneous space  is as in (\ref{so5so3aa}), where in this case
\bea
\mathfrak{so}(3)=\bR\langle T^{(-)}_r\rangle~,~~~\mathfrak{m}=\bR\langle W_a, T^{(+)}_r\rangle~,
\eea
and where $r=1,2,3$ and $a=1,\dots,4$.
Introducing the left-invariant frame as $\ell^A m_A= \ell^a W_a+ \ell^r T_r^{(+)}$, the most general  left-invariant metric is
\bea
ds^2= a \delta_{ab} \bbl^a \bbl^b+ g_{rs} \bbl^r\bbl^s~,
\label{mettrra}
\eea
where  $a>0$ is a constant and  $(g_{rs})$ is any constant $3\times 3$  positive definite symmetric matrix.
The non-vanishing components of the curvature tensor of this metric in the left-invariant frame is
\bea
R_{cd,}{}^a{}_b&=& {a^{-1}\over 16} \Big[ \delta^{ae} \delta^{ pr} \delta^{qs} (I^{(+)}_p)_{ec}  g_{rs} (I^{(+)}_q)_{db}-(d,c)\Big]
-{a^{-1}\over8} \delta^{ae} \delta^{ pr} \delta^{qs}  (I^{(+)}_p)_{eb} g_{rs} (I^{(+)}_q)_{cd}
\cr &&
+{1\over2} (\delta^a_c \delta_{db}- \delta_d^a \delta_{cb})~,
\label{curvsp2sp1aa}
\eea
\bea
R_{rs}{}^a{}_b={a^{-1}\over4} \delta^{pq} g_{pq} \epsilon_{rs}{}^t (I^{(+)}_t)^a{}_b
- {a^{-2}\over 8} \epsilon^{pqt} (I^{(+)}_t)^a{}_b g_{pr} g_{qs}
 -{a^{-1}\over2} \epsilon_{rs}{}^t \delta^{pq} (I^{(+)}_p)^a{}_b  g_{qt}~,
\label{curvsp2sp1ab}
\eea

\bea
R_{ra}{}^s{}_b&=&{1\over8} \big[ g^{sm} \epsilon_{mr}{}^n g_{np}\delta^{pt}+g^{sm} \epsilon_{m}{}^{tn} g_{nr}\big] (I^{(+) }_t)_{ab}
+{1\over8} \epsilon_r{}^{sp} (I^{(+)}_p)_{ab}+ {a^{-1}\over 16} \delta_{ab} \delta^{sm} g_{mr}
\cr
&& +{a^{-1}\over 16} \epsilon^{smn} g_{mr} (I^{(+)}_n)_{ab}~,
\label{curvsp2sp1ac}
\eea
and
\bea
R_{rs, pq}= g_{pl}R_{rs}{}^l{}_q= \epsilon_{rs}{}^m \epsilon_{pq}{}^n  X_{mn}~,
\label{curvsp2sp1ad}
\eea
where
\bea
X_{mn}&=&{1\over2} \delta_{mk} \delta_{nl} g^{kl} (\delta^{q_1q_2} \delta^{p_1p_2} g_{q_1p_1} g_{q_2p_2})
\cr
&&
-2 g_{mn} + \delta_{mn} \delta^{pq} g_{pq}-{1\over4} \delta_{mk} \delta_{nl} g^{kl} (\delta^{q_1q_2} g_{q_1q_2})^2~,
\eea
and the matrix $(g^{rs})$ is the inverse of $(g_{rs})$.
The Ricci tensor in the left-invariant frame is
\bea
&&R_{ab}= -{a^{-1}\over8}  \delta^{pq} g_{pq} \delta_{ab}+{3\over2} \delta_{ab}~,
\cr
&&R_{rs}={1\over4} a^{-2} \delta^{mn} g_{mr} g_{ns}+(\delta_{rs} \delta_{pq} g^{pq} - \delta_{rp} \delta_{sq} g^{pq} ) \delta^{mn} X_{mn}+ \delta_{rp} g^{pm} X_{ms}+\delta_{sp} g^{pm} X_{mr}
\cr &&
- \delta_{pq} g^{pq} X_{rs}-\delta_{rs} g^{pq} X_{pq}~.
\eea
It is straightforward to find the Ricci tensor for $(g_{rs})$ diagonal.
This homogeneous space admits two Einstein metrics one of which is the round sphere metric on $S^7$. This will be explored further in the investigation of the gravitino KSE in section 4.6.3.

\end{document}